\documentclass[acmtog]{acmart}
\usepackage{amsmath}
\usepackage{amsthm}
\usepackage{leftidx}
\usepackage{amssymb}    
\usepackage{subfigure}
\usepackage{soul}
\usepackage{multirow}
\usepackage{xfrac}
\usepackage{mathrsfs}
\usepackage{pdfpages}
\usepackage{color}
\usepackage{afterpage}
\usepackage{subcaption}

\usepackage{diagbox}
\usepackage{algorithm}
\usepackage{algpseudocode}  
\usepackage{enumitem}

\usepackage[normalem]{ulem}
\setcitestyle{square}

\title{Proxy Tracing: Unbiased Reciprocal Estimation for Optimized Sampling in BDPT}

 \author{Fujia Su}
 \authornote{Fujia Su and Bingxuan Li contributed to this work equally.}
 \affiliation{%
   \institution{Peking University}
   \country{China}
 }
 
 \author{Bingxuan Li}
  \authornotemark[1]
 \affiliation{%
   \institution{Peking University}
   \country{China}
 }

 \author{QingYang Yin}
 \affiliation{%
   \institution{Peking University}
   \country{China}
 }
 
 \author{Yanchen Zhang}
 \affiliation{%
   \institution{Peking University}
   \country{China}
 }
 
 \author{Sheng Li}
 \authornote{Sheng Li is the corresponding author.}
 \email{lisheng@pku.edu.cn}
 \affiliation{%
   \institution{School of Computer Science, %National Biomedical Imaging Center, 
   Peking University}
   \country{China}
 }

\begin{document}

\begin{teaserfigure}
  \centering
  \includegraphics[trim={0cm, 4cm, 0cm, 4.8cm},clip, width=1.0\linewidth]{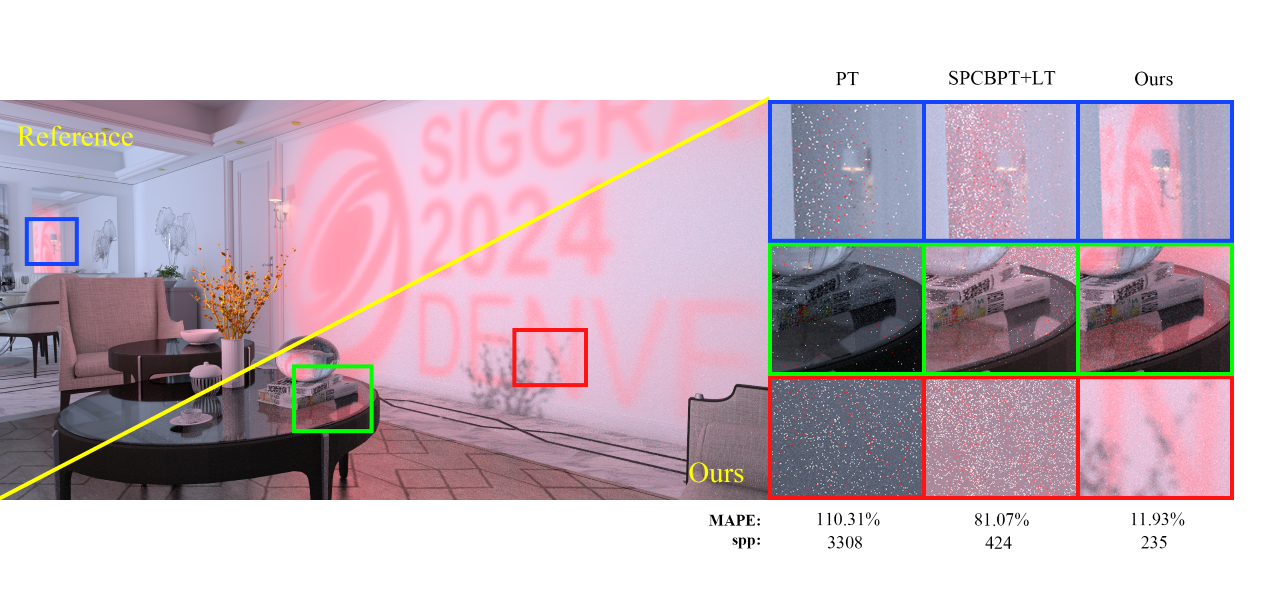}
  \caption{Equal-time ($60s$) comparison between unidirectional Path Tracing (PT) combined with next event estimation (NEE), Subspace-based Probabilistic Connections Bidirectional Path Tracing (SPCBPT) \cite{SPCBPT} with Light Tracing (LT) enabled, and our approach. Mean absolute percentage error (MAPE) is used as the metric for comparison. In this \emph{Projector} scene, a convex lens refracts the textured light from the projector and then projects it on the wall. Our method can efficiently sample the specular-involved difficult paths like $L_{DD}SSDE$ and $L_{DD}SSDSE$, leading to indistinguishable results that closely match the reference. Our approach outperforms the other approaches significantly. }
 \label{fig:teaser}
\end{teaserfigure}

%% Abstract section.
\begin{abstract} 
Robust light transport algorithms, particularly bidirectional path tracing (BDPT), face significant challenges when dealing with specular or highly glossy involved paths. BDPT constructs the full path by connecting sub-paths traced individually from the light source and camera. However, it remains difficult to sample by connecting vertices on specular and glossy surfaces with narrow-lobed BSDF, as it poses severe constraints on sampling in the feasible direction.
To address this issue, we propose a novel approach, called \emph{proxy sampling}, that enables efficient sub-path connection of these challenging paths. When a low-contribution specular/glossy connection occurs, we drop out the problematic neighboring vertex next to this specular/glossy vertex from the original path, then retrace an alternative sub-path as a proxy to complement this incomplete path. This newly constructed complete path ensures that the connection adheres to the constraint of the narrow lobe within the BSDF of the specular/glossy surface.
Unbiased reciprocal estimation is the key to our method to obtain a probability density function (PDF) reciprocal to ensure unbiased rendering. We derive the reciprocal estimation method and provide an efficiency-optimized setting for efficient sampling and connection. 
Our method provides a robust tool for substituting problematic paths with favorable alternatives while ensuring unbiasedness. We validate this approach in the probabilistic connections BDPT for addressing specular-involved difficult paths. Experimental results have proved the effectiveness and efficiency of our approach, showcasing high-performance rendering capabilities across diverse settings.

\end{abstract} % end of abstract

%% Keywords that describe your work. Will show as 'Index Terms' in journal
%% please capitalize first letter and insert punctuation after last keyword

\maketitle

%% ACM Computing Classification System (CCS). 
%% See <http://www.acm.org/class/1998/> for details.
%% The ``\CCScat'' command takes four arguments.

%% Uncomment below to include a teaser figure.

%% Uncomment below to disable the manuscript note
%\renewcommand{\manuscriptnotetxt}{}

%% Copyright space is enabled by default as required by guidelines.
%% It is disabled by the 'review' option or via the following command:
% \nocopyrightspace

%%%%%%%%%%%%%%%%%%%%%%%%%%%%%%%%%%%%%%%%%%%%%%%%%%%%%%%%%%%%%%%%
%%%%%%%%%%%%%%%%%%%%%% START OF THE PAPER %%%%%%%%%%%%%%%%%%%%%%
%%%%%%%%%%%%%%%%%%%%%%%%%%%%%%%%%%%%%%%%%%%%%%%%%%%%%%%%%%%%%%%%%


%\maketitle
%for journal use above \firstsection{..} instead

\section{Introduction} 

\begin{figure}[t]
	\centering
        \begin{minipage}{0.49\linewidth}
            \centering
            \includegraphics[width=0.99\linewidth]{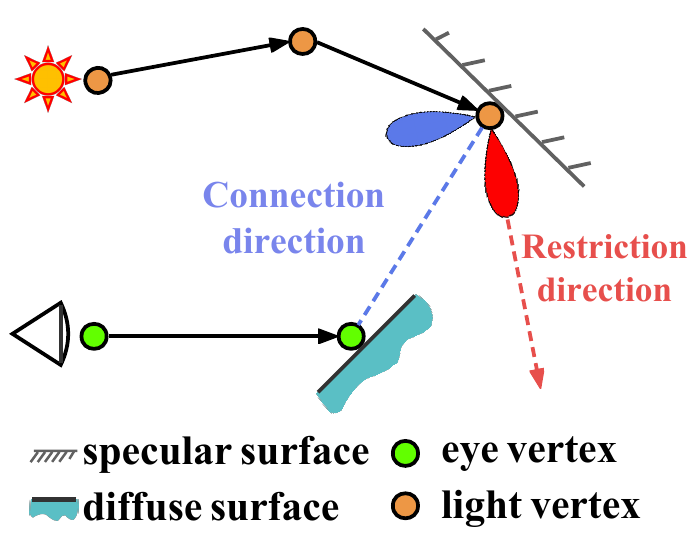}
            \centerline{(a)}
            \label{}
        \end{minipage}
	\centering
        \begin{minipage}{0.49\linewidth}
            \centering
            \includegraphics[width=0.88\linewidth]
            {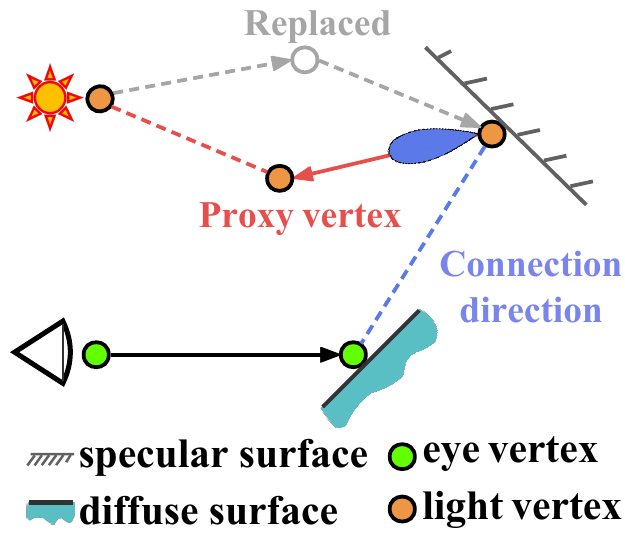}
            \centerline{(b)}
            \label{}
        \end{minipage} 
	\caption{ Motivation of Proxy tracing. (a) When the end vertex of a light sub-path lies on a glossy surface, the probability of connecting it to the eye sub-path within the narrow BSDF lobe is very low. 
 (b) To address this, our method modifies the light sub-path by introducing a proxy vertex instead of the original one, which satisfies the narrow lobe constraint on the glossy surface, thereby enabling us to handle difficult specular-involved paths by improving the connection probabilities.}
	\label{fig:motivation}
\end{figure}

Bidirectional path tracing (BDPT) is a widely used rendering technique known for its robust performance in synthesizing realistic images. 
BDPT employs a variety of sampling strategies to minimize variance, with the majority relying on connection. 
In a connection sampling strategy, an eye sub-path is traced from the camera and a light sub-path is traced from the light source, respectively. These two sub-paths are then connected to construct a full path, and this works well when the end vertices of both sub-paths lie on diffuse surfaces. 
However, when one of the end vertices lies on a specular or highly glossy surface, the connected full path tends to be inefficient, contributing negligibly to the final image.
This is due to the strict constraints required by the Bidirectional Scattering Distribution Function (BSDF) of specular materials with narrow lobe characteristics, which are difficult to satisfy when sub-paths are sampled independently for connections. This issue is exemplified in \autoref{fig:motivation} (a), demonstrating the challenges in connection sampling with specular surfaces.

To improve the efficiency of connection sampling in BDPT, several algorithms based on probabilistic connections have been developed \cite{PCBPT,PCBPT_rismis,SPCBPT}. These algorithms specifically focus on the efficient resampling of light sub-paths given eye sub-path.
However, the challenge of handling high-frequency specular constraints remains. The sampling precision required in such cases is beyond the capabilities of these algorithms.
Another effective technique to reduce variance is Multiple importance sampling (MIS)~\cite{BDPT_mis}, which combines multiple sampling strategies to reduce variance. 
Yet, even with MIS, managing complex paths like $L_{DD}SDSE$\footnote{
We adopt Veach's notation \cite{veach-thesis} to represent paths and light source. Here, $L_{DD},S,D,E$ denote a vertex on the a diffusely emitting light source with finite area, specular surface, diffuse surface, and camera, respectively. Different from Veach's original notation, a simple $E$ is used for the camera because the camera's type is irrelevant in our method. Also, when a surface is highly glossy (metallic/transparent material with roughness smaller than $0.2$ in Disney principled BSDF), we mark the vertex on it as $S$.} remains a formidable task.
In this context, we use a hyphen (-) to indicate connections between the eye sub-path and the light sub-path.
Strategies that involve specular connections, such as $L_{DD}-SDSE$, $L_{DD}S-DSE$, $L_{DD}SD-SE$, and $L_{DD}SDS-E$, exhibit poor performance. 
The only strategy left might involve tracing the whole $L_{DD}SDSE$ path unidirectionally from the camera.
This is usually difficult, particularly when the specular surface $S$ is far from the diffuse surface $D$. 
As a result, none of the existing sampling strategies in BDPT can sample this path efficiently, leading to a significant increase in variance.

Our objective is to tackle challenging paths in BDPT that involve specular or glossy materials with high-frequency BSDF. 
Our key idea is to substitute the less contributive vertices in a light sub-path with newly calculated vertices. This substitution is aimed at creating a more contributive full path, as depicted in \autoref{fig:motivation}.
To accomplish this, we discard the vertex adjacent to the last specular vertex (i.e., the penultimate vertex) in light sub-path, and retrace a new vertex instead, termed the \emph{proxy vertex}. This new vertex is traced based on the direction of the eye sub-path and the BSDF of the specular surface at the light vertex's location. 
By incorporating the proxy vertex into the full path, we ensure adherence to the specular lobe constraints of the BSDF. This adaptation provides a highly effective strategy for sampling these complex paths.

However, when vertices are discarded from a sub-path, it results in an incomplete sub-path which, theoretically, could be originated from an infinite number of initial sub-paths. This presents a significant challenge in unbiasedly estimating its Probability Density Function (PDF), as it requires the integration of all possible original sub-paths. 

Stochastic Multiple Importance Sampling (SMIS) \cite{CMIS} or Marginal Multiple Importance Sampling (MMIS) \cite{MMIS} offer a solution by treating the various original sub-paths as different techniques within a continuous technique space and then applying multiple importance sampling. Nevertheless, SMIS/MMIS face considerable challenges when dealing with the complex sampling distributions that are central to our proxy sampling. Therefore, these approaches are ill-suited for tackling the unique requirements of our methodology.

Instead, we explore an efficient solution based on reciprocal estimation, a method already employed in photon mapping \cite{Qin2015Photon}, volumetric rendering \cite{novak18monte}, specular manifold sampling \cite{Zeltner2020Specular}, and differential rendering \cite{bangaru2020warpedsampling}.
Existing reciprocal estimators are typically constrained by high computational costs and variance. To overcome these limitations, we introduce a novel reciprocal estimator. This new estimator is designed for optimized efficiency and is distinguished by a theoretically derived upper bound on its variance, enhancing its applicability in complex rendering scenarios.

We highlight that our unbiased reciprocal estimator enables the seamless institution of problematic paths with more favorable ones. This replacement is accomplished without compromising the unbiased nature of the Monte Carlo estimation in the path integral framework, and importantly, it is implemented without incurring excessive costs. Thus, our approach enables efficient sampling of the challenging paths using the connection strategy in BDPT.

Finally, our estimator can be integrated with the probabilistic connections BDPT algorithm \cite{PCBPT} to further improve efficiency. We have validated our proxy tracing approach across a range of scene settings through comprehensive experimental comparisons. These comparisons were conducted on both the latest BDPT approaches and sampling approaches, demonstrating the superiority of our approach with high efficiency in handling difficult light transport scenarios. 

Overall, our main contributions can be summarized in two key aspects:

\begin{itemize}
    \item We present a novel path sampling technique using path substitution to greatly enhance BDPT's performance in handling difficult paths. 
    \item We introduce a novel reciprocal estimator. This novel estimator is more efficient and practically applicable, making it a feasible methodology for our path sampling technique.
\end{itemize}

\section{Related Work} 
Our proposed method aims to improve the efficiency of connection sampling strategies in bidirectional path tracing (BDPT), with a focus on difficult paths involving specular materials. While many existing algorithms, including probabilistic connection methods \cite{PCBPT,PCBPT_rismis, SPCBPT}, have been developed for BDPT, our approach can be easily combined with these methods and is specifically designed for efficient sampling of difficult paths.

Caustics, traditionally considered a challenging aspect of specular-involved paths, are commonly rendered using the photon mapping algorithm \cite{PM95}. Conventional photon mapping methods are inherently biased. Progressive photon mapping iteratively evaluates the photon density to produce a biased but consistent rendering result by reducing the kernel radius for radiance estimation \cite{PPM, toshiya2009stochastic}, or by finding a kernel radius that deems the radiance to be unbiased \cite{CPPM, Lin2023TVCG}. \citet{VCM,hachisuka2012ups} propose the VCM/UPS framework to combine BDPT with the photon mapping method. More recently, path guiding strategy has been employed to enable reliable rendering of caustics for path tracing~\cite{li2022unbiased}.

The manifold method \cite{mnee,Zeltner2020Specular} also focuses on specular path handling. By iteratively solving the appropriate position for the specular vertex in a given initial path, thereby producing a path satisfying the specular constraint. This method incorporates reciprocal estimation, as different initial paths may transform into the same final path. Differently, our method fixes the vertex on the specular surface, while the manifold method fixes the neighboring vertices next to the specular vertex. Both the manifold method and reciprocal estimation require multiple iterations and tracing to get an unbiased result. In our method, we retrace only once per connection and significantly reduce the overhead of reciprocal estimation through light sub-path reuse. This efficiency allows our method to integrate seamlessly with existing BDPT methodologies.
In addition, \citet{before_HRR,HRR} proposed the Hierarchical Russian roulette (HRR) and used BVH for connections within the BRDF range, focusing on the SDS path but requiring extensive light sub-path tracing. Our method, conversely, needs fewer light sub-paths and applies to a broader range of path types.

Path guiding algorithms \cite{pathGuiding_lafortune, PG_origin, VarianceAwarePathGuiding} can improve the performance of unidirectional path tracing to better handle the difficult paths during light transport. Path guiding uses the incident radiance distribution to guide path tracing and builds models such as spatio-directional trees \cite{PG_SD} or Gaussian mixture model \cite{PG_GMM} to represent the incident radiance field. Unidirectional path tracing is efficient when the diffuse surface is close to the glossy surface but less so when the surfaces are far apart. Our algorithm, an enhancement of the connection algorithm, performs well, especially when the diffuse surface is distant from the glossy surface. Therefore, our method can be easily incorporated into path-guiding algorithms to provide efficient and robust sampling for difficult paths.

Reciprocal estimation is essential for our method to compute the PDF of proxy tracing. The method to estimate reciprocal unbiasedly was first proposed by \citet{2007Booth} and was introduced to graphics by \citet{Qin2015Photon} to implement unbiased photon gathering. \citet{Zeltner2020Specular} also utilized this method in specular manifold sampling. \citet{Taylor2015} provided the approach to estimate $g(\int x dx)$ unbiasedly by Taylor expansion when function $g$ is nonlinear and smooth. Reciprocal estimation is a special case of the nonlinear smooth function $g$. \citet{debias2022} applied Taylor expansion to evaluate the bias of the biased estimator and transform this estimator into an unbiased one. We adopt the reciprocal estimation framework of the previous method and propose an optimized setting for the reciprocal estimation. Russian roulette and splitting (RRS) method is involved in the reciprocal estimation. \citet{Rath2022EARS} propose the efficiency-optimized setting for the RRS process in path tracing, which inspires us to explore the optimized RRS function in reciprocal estimation.

Traditional MIS methods can only combine the contribution of a certain number of sampling techniques. Continuous MIS (CMIS) \cite{CMIS} extends MIS to an uncountable infinite number of sampling strategies. CMIS utilizes an optimally balanced heuristic to integrate a continuous spectrum of techniques, with its practical counterpart, Stochastic MIS (SMIS), being applicable in most scenes. Marginal MIS (MMIS) \cite{MMIS} further extends this concept to multiple technique spaces, allowing for integration with classical multi-sample MIS estimators. Both SMIS and MMIS can be used to solve our problem, serving as an alternative for reciprocal estimation.

Probabilistic connections methods \cite{SPCBPT, PCBPT, PCBPT_rismis} were developed for resampling the light sub-path based on the information of the eye sub-path. Although these resampling strategies of the light sub-path may not be sufficient to satisfy specular constraint, they can be combined with our method to cover the sampling of various difficult paths. By connecting sub-paths and generating valid specular paths, we can overcome difficulties associated with sampling specular materials, as illustrated in \autoref{fig:motivation}(b). We also adopt the idea of light sub-path reuse in the probabilistic connections method to reduce the overhead of reciprocal estimation; as well as the idea of selecting the appropriate light sub-path for connection to ensure the best possible efficiency and rendering quality.

\section{PRELIMINARIES}
\subsection{Path Integral and Bidirectional Path Tracing}
From the path integral formulation~\cite{BDPT_original}, the pixel measurement $I$ is as:
\begin{equation}
\label{equation:path intergral}
	I = \int _{\Omega}f(\bar{x})d\mu(\bar{x}) \ ,
\end{equation}
where $\Omega$ is the path space, and $\bar{x} = x_{0}x_{1}...x_{k}$ is a path of length $k \geq 1$,  $x_{0}$ and $x_{k}$ are on a light source and the camera, respectively. $d\mu(\bar{x})=dA(x_{0})...dA(x_{k})$ is the differential area product, and $f$ is the measurement contribution function as:
\begin{gather*}
f(\bar{x})=L_{e}(x_{0},x_{1})T(\bar{x})W_{e}(x_{k-1},x_{k}) \ ,\\
T(\bar{x}) = GV(x_{0},x_{1})\left [ \prod _{i=1}^{k-1}\rho (x_{i-1},x_{i},x_{i+1})GV(x_{i},x_{i+1}) \right ] \ ,
\end{gather*}
where $L_{e}(x_{0},x_{1})$ is the radiance emitted $x_{0}$ $\rightarrow$ $x_{1}$, $W_{e}$ is the pixel sensitivity, $\rho$ is the bidirectional scattering distribution function (BSDF), and $GV$ is the geometry term including the visibility.
 
BDPT estimates \autoref{equation:path intergral} by Monte Carlo integration. It traces the light sub-path $\bar{y}$ from the light source in PDF $p(\bar{y})$ and eye sub-path $\bar{z}$ from the camera in PDF $p(\bar{z})$, then connects these two sub-paths to sample the full path $\bar{x}=\bar{y}\bar{z}$. The PDF for sampling $\bar{x}$ is $p(\bar{x}) = p(\bar{y})p(\bar{z})$. The contribution of the full path $f(\bar{x})$ can be divided into three parts $f_{z}(\bar{z})f_{yz}(\bar{y},\bar{z})f_{y}(\bar{y})$ as~\cite{PCBPT}:
\begin{equation*}
\begin{aligned}
\label{equ:light_eye_connection}
f_y(\bar{y}) = L_e(y_0,y_1)GV(y_0,y_1)\prod _{i=1}^{s-2}\rho(y_{i-1},y_i,y_{i+1})GV(y_{i},y_{i+1}), \\
f_{yz}(\bar{y},\bar{z}) = \rho(y_{s-2},y_{s-1},z_{t-1})GV(y_{s-1},z_{t-1})\rho(y_{s-1},z_{t-1},z_{t-2}),\\
f_z(\bar{z}) = W_{e}(z_0,z_1)GV(z_0,z_1)\prod _{i=1}^{t-2}\rho(z_{i-1},z_i,z_{i+1})GV(z_{i},z_{i+1})\ .
\end{aligned}
\end{equation*} 

Here, $s$ is the vertices number of the light sub-path, and $t$ is the vertices number of the eye sub-path. 

The PDF for a sub-path to be traced $p(\bar{y})$ can be written as the PDF product of its vertices as
\begin{equation} p(\bar{y})=p(y_{0})p(y_1)....p(y_{s-1})
\end{equation}
and the PDF for a vertex $p(y_{i})$ to be traced usually depends on its precedent vertex $p(y_{i-1})$ and the incident direction $y_{i-2}\rightarrow y_{i-1}$ to trace from BSDF at $y_{i-1}$
\begin{equation}
\label{equ:tracingPdf}
    p(y_{i}) = p(y_{i-2}\rightarrow y_{i-1}\rightarrow y_{i}) = p(y_{i}|y_{i-1},y_{i-2}\rightarrow y_{i-1}) \ .
\end{equation}

\subsection{Probabilistic Connections}
The probabilistic connections method, an enhancement of BDPT, boosts sub-path connection efficiency \cite{PCBPT}. It begins each iteration by caching a set of light sub-paths. These are then resampled when connecting with the eye sub-paths, improving efficiency by reusing the cached light sub-paths for all eye sub-paths and selectively providing higher sampling density to paths with greater contributions.

SPCBPT introduces the concept of subspace within path space to implement efficient probabilistic connections \cite{SPCBPT}. Sub-paths are classified into subspaces based on position, normal, and incident direction, ensuring low discrepancy and shared sampling probabilities within each subspace. The method employs a two-stage process for light sub-path selection: first, selecting a light subspace based on the eye subspace, followed by resampling the light sub-path from this chosen subspace. In probabilistic algorithms like SPCBPT, $p(\bar{y})$ and $p(\bar{z})$ can easily approximate the shape of $f_y(\bar{y})$ and $f_z(\bar{z})$ by tracing from BSDF. In most cases, the probabilistic connections method can provide good sampling for $f_{yz}$. However, when $y_{s-1}$ or $x_{z-1}$ is on specular surface, the high variance of $f_{yz}(\bar{y},\bar{z})$ will make the Monte Carlo estimator $\hat{I} = f/p = \frac{f_y(\bar{y})}{p(\bar{y})}\frac{f_{z}(\bar{z})}{p(\bar{z})}f_{yz}(\bar{y},\bar{z})$ less efficient. Even the probabilistic connection method can hardly find the appropriate light sub-path that satisfies the high-frequency specular constraint. 

\begin{figure*}[t]
	\centering
        \begin{minipage}{0.3\linewidth}
            \centering
            \includegraphics[width=0.9\linewidth]            {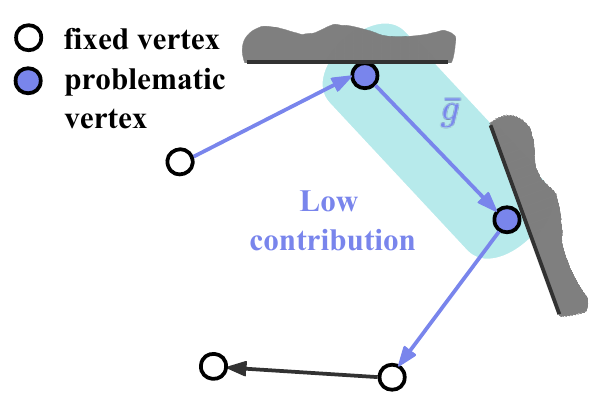}
            \centerline{(a)}
            \label{}
        \end{minipage}
	\centering
        \begin{minipage}{0.3\linewidth}
            \centering
            \includegraphics[width=0.9\linewidth]
            {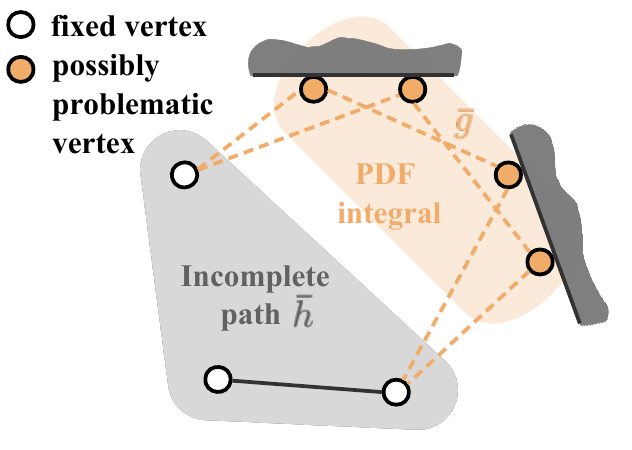}
            \centerline{(b)}
            \label{fig:proxy_sampling_b}
        \end{minipage} 
        \centering
        \begin{minipage}{0.3\linewidth}
            \centering
            \includegraphics[width=0.9\linewidth]
            {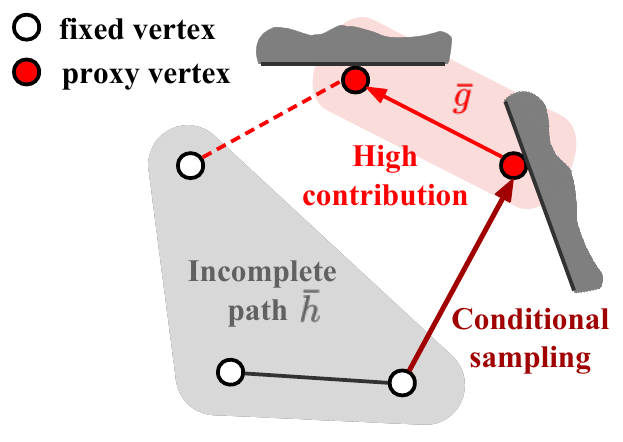}
            \centerline{(c)}
            \label{}
        \end{minipage} 
	\caption{Semantic illustration of proxy sampling. (a) An original path, sampled from the initial distribution, often includes poorly distributed vertices, reducing the overall contribution of the full path; (b) our approach discards the problematic vertices, resulting in an incomplete path. The PDF for an incomplete path requires integrating over all potential problematic paths that could lead to the identical incomplete path; (c) proxy vertices are then traced to merge with the incomplete path and construct a new path. The data from the incomplete sub-path assists in tracing these proxy vertices, providing a more favorable conditional distribution for constructing a full path with a higher contribution.}
	\label{fig:proxy_sampling}
\end{figure*}

Although multiple importance sampling (MIS) can leverage multiple sampling strategies in BDPT and select an appropriate strategy for the specular material \cite{CMIS, kondapaneni2019optimal}, the fundamental challenges of path sampling and connection are still not fully addressed. Sampling paths involving specular materials, such as $L_{DD}(S|D)^{\ast} SDS^{\ast}E$ remains difficult.
In this type of path, all the possible sampling strategies rely on the eye sub-path to trace from its first diffuse surface to the next specular surface and the PDF for the $D\rightarrow S$ bounce can be very small. Consequently, it is challenging for BDPT to sample such path effectively, which motivates the development of our proposed method. % a connection is unavailable in the $S^{\ast}$ segment before the diffuse vertex, 

Constructing a path with specular connections, like 
$L_{DD}(S|D)^{\ast}S$-$DS^{\ast}E$, often fails to satisfy specular constraint, thus yielding negligible contributions to the final image. To address this issue, we propose retracing the neighbor of the specular vertex that can satisfy specular constraint, converting the invalid path into valuable ones with high contribution.
In \autoref{section:principle}, we will introduce our methodology of proxy sampling. This technique involves discarding the low-contribution segment of the path and retracing the complementary proxy path to enhance sampling performance. 

\subsection{Continuous, Stochastic and Marginal MIS}
\label{sec:CMIS-SMIS-MMIS}

Multiple Importance Sampling (MIS) \cite{BDPT_mis} is a Monte Carlo integration framework for combining various sampling techniques, yet traditionally limited to countable sets. Continuous MIS (CMIS) \cite{CMIS} extends this to an uncountable technique space $\mathcal{T}$, broadening its applicability. The one-sample estimator in CMIS is 
\begin{equation}
\label{equ:CMIS_estimator}
    \scalebox{0.8}[1.5]{<}I\scalebox{0.8}[1.5]{>}_{\mathrm{CMIS}}=\frac{w(t,x)f(x)}{p(t)p(x|t)},
\end{equation}
where weighting function $w: \mathcal T \times \mathcal X \to \mathbb R$ with the property $\int_\mathcal{T}w(t, x)\mathrm{d}t = 1$. For the estimator to be unbiased, the weighting function $w$ must satisfy
\begin{subequations}
\begin{align}    
    \int_{\mathcal{T}}w(t,x)dt=1\ \mathrm{whenever}\ f(x)\neq0, \\
    w(t,x)=0\ \mathrm{whenever}\ p(t,x)=0.
\end{align}
\end{subequations}

A variance-optimal weighting function, as derived in \cite{CMIS}, is
\begin{equation}
\label{equ:CMIS_weight_integral}
  \tilde{w}(t,x)=\frac{p(t)p(x|t)}{\int_{\mathcal{T}}p(t',x)\mathrm{d}t'}=\frac{p(t,x)}{p(x)}, 
\end{equation}
apply this optimal weighting function in \autoref{equ:CMIS_estimator}, we get the balance-heuristic CMIS estimator
\begin{equation}
\label{equ:balance CMIS definition}
\scalebox{0.8}[1.5]{<}I\scalebox{0.8}[1.5]{>}_{\mathrm{CMIS}}=\frac{\tilde{w}(t,x)f(x)}{p(t,x)}=\frac{f(x)}{p(x)}.
\end{equation}

Evaluating the balance-heuristic CMIS estimator requires evaluating the marginal PDF integral $p(x)$ in its denominator. Typically, $p(x)$ is not readily available in a closed form. Our solution is to create an unbiased estimator for the reciprocal, $1/p(x)$. \cite{CMIS} circumvented reciprocal estimation by proposing Stochastic MIS (SMIS), which can be seen as an approximation for balance-heuristic CMIS.

In SMIS, $n$ independent sample pairs $(t_1, x_1), . . ., (t_n, x_n)$ are sampled. Then the SMIS estimator is constructed as
\begin{equation}
\scalebox{0.8}[1.5]{<}I\scalebox{0.8}[1.5]{>}_{\mathrm{SMIS}}=\sum^{n}_{i=1}\frac{\dot{w}(t_i,x_i)f(x_{i})}{p(x_i|t_i)}=\sum^{n}_{i=1}\frac{f(x_i)}{\sum^{n}_{j=1}p(x_i|t_j)},
\end{equation}
where $\dot{w}(t_i,x)=\frac{p(x|t_i)}{\sum^n_{j=1}p(x|t_j)}$. 
SMIS effectively samples techniques from the space $\mathcal{T}$ and then applying MIS.
SMIS exhibits bias when there’s a positive probability that the combined sampling distribution cannot cover $f(x)$’s support. Therefore, it requires nearly every technique in $\mathcal{T}$ being unbiased, i.e. $p(x|t)>0$ whenever $f(x)>0$, if the techniques are chosen in a stochastic way. 
Further challenges arise when each technique only provide efficient sampling for small range of $x$. In that case, SMIS requires an large number of techniques and appropriate importance sampling in 
$\mathcal{T}$, both can be difficult to implement. The computational cost, scaling as 
$O(n^2)$, further limits the number of combinable techniques. 
We conclude these factors as the main limitations of SMIS.

Marginal MIS (MMIS) \cite{MMIS} generalized SMIS to multiple technique spaces. 
While MMIS enhances the robustness of SMIS by incorporating multiple technique spaces, it still inherits the limitations of SMIS. If none of the technique spaces within MMIS are capable of efficiently sampling  $f(x)$, then the MMIS estimator can not achieve high efficiency, as it just combines these spaces.

These identified limitations of SMIS/MMIS underpin our decision to favor reciprocal estimation in our approach. We will provide further illustration in \autoref{sec:comparison with CMIS}, where we conduct a comparison between SMIS/MMIS and our reciprocal estimation. 

\section{Our Methodology} 
\label{section:principle} 

When sampling a path $\bar{x}$ from a distribution $p(\bar{x})$, certain vertices in $\bar{x}$ may contribute to high variance due to poor distribution. To mitigate this, we employ a dropout-and-complement scheme, removing these \emph{problematic} vertices and retracing new ones as substitutes. The retracing process is designed to be independent of the original problematic vertices while leveraging information from the fixed vertices in the original path. 
In this section, we formulate this dropout-and-complement scheme and show that its core principle in a reciprocal estimation of PDF for unbiasedness, and is effectively addressed by our proposed estimator eventually.

\subsection{Proxy Sampling Formulation}
\label{sec:proxy sampling formulation}
Formally, we use $\bar{g}$ to denote the candidate vertices that can be replaced in path $\bar{x}$. $\bar{h}$ to denote the incomplete path that contains fixed vertices which will be used in the retracing. We use function $D(\bar{x},\bar{g})$ to denote the operation of dropping out problematic vertices in $\bar{g}$ from complete $\bar{x}$, obtaining an incomplete path $\bar{h}$:
\begin{equation}
    \bar{h} = D(\bar{x},\bar{g}).
\end{equation}
We also define the repair operation $R(\bar{h},\bar{g})$ to fill in this incomplete path $\bar{h}$ with proxy vertices $\bar{g}$ and get the complete path $\bar{x}$:
\begin{equation}
    \bar{x}=R(\bar{h},\bar{g}).
\end{equation}

We illustrate the key aspects of proxy sampling in \autoref{fig:proxy_sampling}. We first apply dropout operation $D(\bar{x}_{o},\bar{g}_{o})$ to obtain an incomplete path $\bar{h}$ from original path $\bar{x}_{o}$ (see \autoref{fig:proxy_sampling}(a)). Next, we retrace proxy vertices $\bar{g}$ from distribution $p(\bar{g}|\bar{h})$ and generate the final path by applying the repair operation $\bar{x} = R(\bar{h},\bar{g})$ (see \autoref{fig:proxy_sampling}(c)). Hereby, The PDF for sampling the final path $p_{d}(\bar{x})$ is given by
\begin{equation}
    p_{d}(\bar{x}) = p(\bar{h})p(\bar{g}|\bar{h}),
\end{equation}
where $p(\bar{h})$ is used instead of $p(\bar{x}_{o})$ to compute the PDF. As shown in \autoref{fig:proxy_sampling}(b), an incomplete path $\bar{h}$ can result from different original paths with different alternative vertices $\bar{g}$, so the PDF of incomplete path $\bar{h}$ should be computed as 
\begin{equation}
\label{equ:incomplete_integral}
    p(\bar{h}) = \int_{A^{u}} p[R(\bar{h},\bar{g})]d\mu(\bar{g}),
\end{equation}
where $A$ is the surfaces of a scene and $u$ is the vertex count (also path length) of $\bar{g}$.
 
Proxy sampling can provide better performance than the original distribution $p(\bar{x})$ when the conditional distribution $p(\bar{g}|\bar{h})$ is efficient. 
For instance, when a specular vertex and one of its neighbor vertices are located within $\bar{h}$, the conditional distribution for sampling the other neighbor vertex in $\bar{g}$ is efficient by sampling from the specular BSDF.

However, computing the PDF $p(\bar{h})$ for sampling an incomplete path, as given in \autoref{equ:incomplete_integral}, requires an integral that cannot be computed analytically in most cases. We can get an unbiased estimator $\tilde{p}(\bar{h})$ of $p(\bar{h})$ by applying the Monte Carlo method to \autoref{equ:incomplete_integral}. Unfortunately, $\frac{1}{\tilde{p}(\bar{h})}$ is not an unbiased estimator for $\frac{1}{p(\bar{h})}$ \cite{1987Statistic_Introduction}. This means the path integral estimator $f(\bar{x})/[\tilde{p}(\bar{h})p(\bar{g}|\bar{h})]$ is biased due to the presence of $\tilde{p}(\bar{h})$ in the denominator. Therefore, we need an unbiased reciprocal estimation for $\frac{1}{p(\bar{h})}$ in the proxy sampling. By deriving a reciprocal estimation formulation for the path integral, we can effectively estimate the PDF of the incomplete paths without introducing bias.

\subsection{Our Reciprocal Estimation} 
\label{sec:reciprocalEstimation}
In the reciprocal estimation, we adopt the Taylor expansion framework of \citet{Taylor2015}. Without loss of generality, we use $\beta$ to denote the integral of a function $\beta = \int_{\mathcal{R}} f(x)dx$ in an integral domain $\mathcal{R}$. We use a supporting distribution $q$ to sample $x$ from $\mathcal{R}$. Our objective is to get an unbiased estimation for the reciprocal of $\beta$ as
 $$\alpha = \frac{1}{\beta} = \frac{1}{\int_{\mathcal{R}}{ f(x)} dx}.$$

We first compute the difference between $\beta$ and a constant $B$ as $A = B - \beta$, then we have $$\alpha = \frac{1}{B-A}.$$
A Taylor expansion of $\frac{1}{B-A}$ at $A=0$ yields
\begin{equation}
\label{equ:taylor}
   \frac{1}{B-A} = \frac{1}{B}+\frac{A}{B^2}+\frac{A^2}{B^3}...=\frac{1}{B}\sum_{n=0} \frac{A^n}{B^n}.  
\end{equation}
 
The integral formulation of \autoref{equ:taylor} is
\begin{equation}
    \label{equ:reciprocalIntegral_0}
\alpha = \frac{1}{B}\sum_{n=0} I_n=\frac{1}{B}\sum_{n=0} \int_{\mathcal{R}^n}\prod_{i} \frac{B-f(x_i)}{B}d\mu(\bar{x}),
\end{equation}
where $I_n = \frac{A^n}{B^n}$ and $\bar{x}$ is constructed by $N$ samples from domain $\mathcal{R}$.

The product of $n$ random variables with $\frac{A}{B}$-expectation is an unbiased estimation for $\frac{A^n}{B^n}$. Therefore, we can estimate $I_{n}$ without bias by generating $n$ independent samples from the supporting distribution $q$ and building estimator $\tilde{I}_{n}$ as 
\begin{equation}
    \label{equ:reciprocalEstimator_n}
    \tilde{I}_{n}= \prod_{i}^{n} \left [ 1 - \frac{f(x_i)}{Bq(x_i)}\right ] .
\end{equation}
Here, $E[1-\frac{f(x_i)}{Bq(x_i)}] = \frac{A}{B}$, so $\tilde{I}_{n}$ is a unbiased estimator for $I_n = \frac{A^n}{B^n}$. We use $1-\frac{f(x_i)}{Bq(x_i)}$ instead of $\frac{B-f(x_i)}{Bq(x_i)}$ to construct the $\frac{A}{B}$-expectation random variable because the optimal supporting distribution $q$ for the former is $q \propto f(x)$ while the latter is $q\propto 1-f(x)$, which is typically more challenging and unconventional to construct in most scenarios.

We can make an unbiased estimation to $\alpha$ by combining 
infinite $\tilde{I}_{n}$ as
\begin{equation}
\label{equ:reciprocalEstimator0}
\tilde{I} = \frac{1}{B} \sum_n \tilde{I}_n .
\end{equation}
\citet{Taylor2015} proposes two solutions to address the infinite summation in \autoref{equ:reciprocalEstimator0}. One approach constructs a single-term estimator: Sample an integer \( N \) from \( \mathbb{Z} \) with a probability mass function (PMF), \( p_N \), generate \( N \) samples to estimate \( \tilde{I}_N \), and use \( \tilde{I}_N / [Bp_{N}] \) as the unbiased estimator for \( \tilde{I} \). The other also samples \( N \) and generates \( N \) samples, but instead of estimating \( \tilde{I}_{N} \), it uses the \( N \) samples to estimate any \( \tilde{I}_{N'} \) for \( N' \le N \) by computing the prefix product. The resulting estimator for \( \tilde{I} \) is then given by \( \sum_{i \le N}\tilde{I}_{i} / [Bp(j \ge i)] \). Both solutions rely on appropriate distribution to sample integer $N$. Although \citet{Taylor2015} introduced a cost-minimized distribution, applying this distribution to our proxy tracing method still presents challenges. This arises from several non-closed terms required by the distribution, such as the root of a complex non-linear equation, which are impractical to solve for our estimation. Consequently, we have chosen to deviate from their methodology and instead focus on addressing the reciprocal estimation problem with a simpler and more practical solution.

Since \autoref{equ:reciprocalIntegral_0} is close to the path integral formulation in \autoref{equation:path intergral}, it can be estimated in a Russian Roulette and Splitting (RRS) manner, similar to unidirectional path tracing. This involves constructing $\bar{x}$ from the first sample $x_{0}$ and controlling the sampling using an RRS function $r(\bar{x})$. When a prefix $\bar{x}$ is sampled and $R(\bar{x})\le 1$, we finish the sampling in probability $1-r(\bar{x})$; otherwise, continue the sampling. If $r(\bar{x})\ge 1$, we split $\left \lfloor r(x) \right \rfloor + 1$ samplings on the prefix $\bar{x}$ in probability $r(x) - \left \lfloor r(x) \right \rfloor$; otherwise, split $\left \lfloor r(x) \right \rfloor $ samplings. The estimator for a path of $k$ vertices $\bar{x}$ is 
\begin{equation}
    \label{equ:RRS_estimator}
     \tilde{I}(\bar{x}) = \frac{\prod_{0\le i<k} \left [ 1 - \frac{f(x_i)}{Bq(x_i)}\right ] }{\prod_{1\le j< k}r(\bar{x}_{j})},
\end{equation}
where $\bar{x}_{j}$ is the $j$-vertex prefix of $\bar{x}$. $ \tilde{I} = \left [1 + \sum \tilde{I}(\bar{x})\right]/B$ provides unbiased estimation to $\alpha$, where the extra $\frac{1}{B}$ comes from $\tilde{I}_{0}$.

We propose an RRS function for a $\bar{x}$ of $k$ vertices as
\begin{equation}
    \label{equ:rr}
    r(\bar{x}) = |1 - \frac{f(x_{k-1})}{Bq(x_{k-1})}|.
\end{equation}
We prove that the proposal RRS function is sub-optimal, i.e., optimal in some assumptions. We provide a detailed explanation of our estimator with analysis and proof in \autoref{appendix:estimator}.
With our proposal RRS function, estimator $\tilde{I}(\bar{x})$ can be transformed into
\begin{equation} 
\label{equ:final_estimator}
\tilde{I}(\bar{x}) = \left[  1 - \frac{f(x_{k-1})}{Bq(x_{k-1})}\right] S(\bar{x}),
\end{equation}
where $S(\bar{x}) = \prod_{0 \le j < k-1} sgn[1 - \frac{f(x_{j})}{Bq(x_{j})}]$ is used to denote the sign product of all the prefix vertices of $\bar{x}$. 

In \autoref{sec:optimal value for B}, we also discuss the optimal setting of $B$. With our proposed RRS function, we construct a sign estimator $\tilde{I}_{sign} (\bar{x})= S(\bar{x})$ which can also be used for reciprocal estimation but produces higher variance than $\tilde{I}(\bar{x})$. We provide the variance upper bound as well as the work-minimized B setting for the sign estimator. The recommended $B$ setting is $B = max(\frac{f(x)}{q(x)})$, i.e., the upper bound of $\frac{f(x)}{q(x)}$. We use the upper bound $B$ setting for our estimator. Because $V(\tilde{I})\le V(\tilde{I}_{sign})$, we highlight that using this $B$ can minimize the upper bound of reciprocal estimation.

In contrast to \citet{Taylor2015}, our RRS method does not directly sample $N$. Instead, it determines the number of samples by continuous application of RRS on current prefix samples. The probability of sampling $\bar{x}$ is decided by both $\bar{x}$'s length and the prefix's importance (production of $r(x_i)$). Note that \citet{2007Booth} uses a similar Russian roulette method in reciprocal estimation (splitting not involved). However, the RR function used is $r(\bar{x}) = min(1, \frac{1}{r}\prod_{i<k} |1-\frac{f(x_i)}{Bq(x_i)}|)$ where $r$ is a hyperparameter that must be manually set. In contrast, our Russian roulette method uses a sub-optimal RRS function that is supported by rigious variance analysis, without any manually-set parameters required.
Additionally, both our method and Booth's method can be adapted to the geometric distribution when dealing with binary functions $f(x)$, as demonstrated by \citet{Qin2015Photon}.

\begin{figure}[t]
	\centering
 \includegraphics[trim={1cm, 1.5cm, 0cm, 0cm},clip,width=0.7\linewidth]{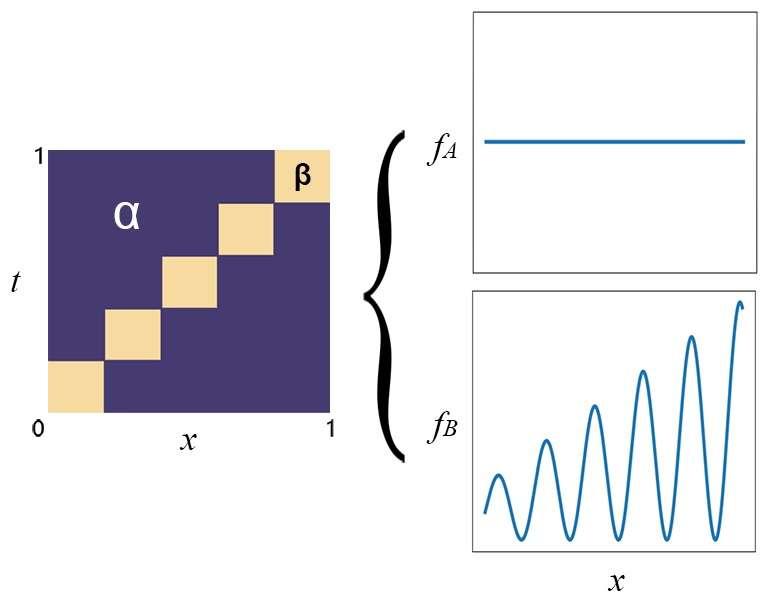}
 \caption{
Illustration of the sampling distribution $p(t,x)$ alongside the target function $f(x)$. Parameters $\alpha$ and $\beta$ in the graph correspond to the PDF in this region. Different $t$ associates with different sampling techniques $p(x|t)$. SMIS works by first sampling $n$ techniques $t_1,t_2,...,t_n$, then performing MIS between $p(x|t_1),p(x|t_2),...,p(x|t_n)$.}
 \label{fig:smis figure 2}
\end{figure}

\begin{table}[thb]
\centering	
\caption{
Results for $\alpha=0.1$ and $\beta=4.6$. We compared the variances of SMIS-$n$ and our reciprocal estimator under conditions of equal cost. SMIS16 shows the best performance within this specific setting.}

\label{tab:SMIS2-tab1}       % Give a unique label
\begin{tabular}{c|c|c|c|c|c}
\hline
 {\diagbox[]{}{}} & SMIS2 & SMIS4 & SMIS8 & SMIS16  & OURS \\
\hline
$f_A$ & 22.0 & 15.2 & 7.5 & \textbf{2.1} & 4.3 \\
\hline
$f_B$ & 203.5 & 142.4 & 71.6 & \textbf{23.8} & 51.7 \\
\hline

\end{tabular}
\end{table}
\begin{table}[thb]
\centering	
\caption{
Results for $\alpha=0.01$ and $\beta=4.96$. Our reciprocal estimator exhibited superior efficiency. Note that SMIS-$n$'s variance increases dramatically compared to \autoref{tab:SMIS2-tab1} while our estimator maintains nearly consistent variance across tests, showing our method's robustness in handling this type of distribution. }
\label{tab:SMIS2-tab2}       % Give a unique label
\begin{tabular}{c|c|c|c|c|c}
\hline
 {\diagbox[]{}{}} & SMIS2 & SMIS4 & SMIS8 & SMIS16  & OURS \\
\hline
$f_A$ & 240.2 & 173.6 & 91.0 & 21.6 & \textbf{4.5} \\
\hline
$f_B$ & 2344.8 & 1658.4 & 874.9 & 197.1 & \textbf{54.0} \\
\hline
\end{tabular}
\end{table}

\subsection{Comparative Analysis of SMIS/MMIS and our Reciprocal Estimation}
\label{sec:comparison with CMIS}
In this section, we will demonstrate the conditions under which our reciprocal estimation method outperforms SMIS/MMIS, and explain our preference for the former. Our analysis is structured in two parts: Initially, we will conduct unit experiments to compare the performance of each method directly. Subsequently, we will explore complex scenarios where SMIS/MMIS exhibit limitations, thereby emphasizing the strengths of our proxy sampling approach.

\subsubsection{Unit Test}

Our unit test is depicted in \autoref{fig:smis figure 2}, where we employ a sampling distribution $p(t,x)$ characterized by parameters $\alpha$ and $\beta$. These parameters are adjustable to estimate $f(x)$. It is important to note that only one parameter can be varied at a time, as they must satisfy the condition $0.8\alpha + 0.2\beta = 1$ to ensure the validity of the distribution.

\autoref{tab:SMIS2-tab1} and \autoref{tab:SMIS2-tab2} show the variance of SMIS-$n$ and our reciprocal estimator under conditions of equal cost. In \autoref{tab:SMIS2-tab1} where $\alpha=0.1$, SMIS16 performs best. However, as $\alpha$ is reduced to 0.01 in \autoref{tab:SMIS2-tab2}, our reciprocal estimator exhibited superior efficiency. The variance of SMIS-$n$ increases dramatically as $\alpha \to 0$. This phenomenon can be attributed to the characteristics of the distribution $p(t,x)$. As $\alpha\to 0$, $p(x|t)$ associate with different $t$ targets different segments of $f(x)$. Notably, each individual sampling technique $p(t,x)$ is characterized by significant variance, and the combination of a limited subset fails to mitigate this issue. Only by combining all five distinct sampling methods (see \autoref{fig:smis figure 2}) can SMIS get comprehensive coverage of $f(x)$'s support.

When $\alpha=0$, this situation falls outside SMIS's scope, and SMIS will produce bias. This bias arises because a combination of different sampling techniques does not consistently cover the entire support of $f(x)$. While increasing the number of techniques used in SMIS may reduce this bias, a certain level of positive bias remains inevitable as long as the number of techniques is finite.

Although MMIS was not included in the direct experimental comparison, it is crucial to recognize that MMIS shares the same limitations as SMIS. This similarity arises because MMIS operates by integrating multiple SMIS strategies.

\subsubsection{SMIS/MMIS in Difficult Situations}
\begin{figure}[tb]
     \centering
     \centering
    \subfigure[$L_{DD}SS$ path]{
    \includegraphics[trim={0.2cm, 0.1cm, 0.0cm, 0.1cm},clip,width=0.4\linewidth]{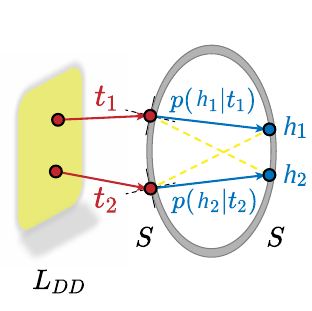}
     }\hspace{-4mm}
    \subfigure[Visibility discrepancy]{
    \includegraphics[trim={0cm, 0cm, 0cm, 0cm},clip,width=0.6\linewidth]
    {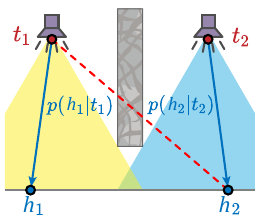}
     }\hspace{-4mm}
\caption{
Two failure cases in SMIS. (a) In handling this $L_{DD}SS$ type light sub-path, our proxy sampling designates the $L_{DD}S$ as the dropout vertices, represented by $\bar{g}_o$, while the residual $S$ forms the incomplete path $\bar{h}$. Within SMIS framework, $\bar{g}_o$ is associated with technique $t$, and $p(\bar{h}|t)$ is proportional to the PDF that traces from $\bar{g}_o$'s last vertex to the first vertex (denote as $h$ without the bar) following $\bar{g}_o$ in $\bar{h}$. The probabilities $p(h_2|t_1)$ and $p(h_1|t_2)$, signified by yellow dashed lines, are nearly zero. 
(b) In scenario where visibility varies significantly, different techniques in SMIS primarily sample separate regions of \(f(x)\). This leads to an inherent bias in SMIS because there is a positive probability that the combined sampling distribution cannot cover $f(x)$'s support.
}
\label{fig:smis-proxy-1}
\hspace{-5mm}
\end{figure}

Our proxy sampling is designed to handle specular-involved complex situations, while SMIS/MMIS will face challenges in such contexts.
Recall Sec. \ref{sec:proxy sampling formulation}, we first trace the original path $\bar{x}_{o}$ from the original sampling method, and "dropout" vertices $\bar{g}_o$ from the original path $\bar{x}_{o}$ to get the incomplete path $\bar{h}$, and complement $\bar{h}$ with newly traced proxy vertices $\bar{g}$. The "dropout" vertices $\bar{g}_{o}$ is actually the technique in CMIS context, since it determines the sampling of $\bar{h}$. Also, $A^u$ is the technique space. In the section, we use notations $t = \bar{g}_{o}$ and $\mathcal{T} = A^u$ to fit in with the CMIS context. The balance-heuristic CMIS estimator is then constructed as

\begin{equation}
\label{equ:CMIS in proxy tracing}
 \scalebox{0.8}[1.5]{<}I\scalebox{0.8}[1.5]{>}_{\mathrm{CMIS}}=\frac{f(\bar{x})}{p(\bar{x})} =\frac{f(\bar{x})}{p(\bar{h})p(\bar{x}|\bar{h})}=\frac{1}{\int_{\mathcal{T}}p(\bar{h},t)dt} {\cdot} \frac{f(\bar{x})}{p(\bar{x}|\bar{h})}
 .
\end{equation}
In SMIS, \(n\) independent pairs of techniques and corresponding samples, denoted as \((t_1, \bar{x}_1), \ldots, (t_n, \bar{x}_n)\), are drawn. The SMIS estimator is constructed as
\begin{equation}
      \scalebox{0.8}[1.5]{<}I\scalebox{0.8}[1.5]{>}_{\mathrm{SMIS}}= \sum^{n}_{i=1}\frac{1}{\sum^{n}_{j=1}p(\bar{h}_i| t_j)}{\cdot}\frac{f(\bar{x}_i)}{p(\bar{x}_i|\bar{h}_i)}.
\end{equation}
Here, $p(\bar{x}_i|\bar{h}_i) = p(\bar{g_i}|\bar{h}_{i})$ is the retracing PDF; $p(\bar{h}|t)$ refers to the PDF for all the vertices in $\bar{h}$ to be traced in the original sampling method when the vertices in $t$ given. Note that  computing $p(\bar{h}_i|t_j)$ requires a visibility test between $\bar{h}_i$ and $t_j$ when $i\neq j$, as illustrated in \autoref{fig:smis-proxy-1}. This requirement results in a quadratic increase in the cost of visibility tests within SMIS. Consequently, utilizing smaller numbers of techniques is generally more practical in proxy tracing.

We present two scenarios illustrating the lesser effectiveness of SMIS. The first involves multiple specular vertices following a light vertex, as depicted in \autoref{fig:smis-proxy-1} (a). Constraints in reflective or refractive directions result in \(p(h_2|t_1)\) and \(p(h_1|t_2)\) being almost zero. This associates with \autoref{fig:smis figure 2} when $\alpha \to 0$, where different techniques predominantly sample different regions of \(f(x)\). In cases where the \(S\) vertex is highly glossy, SMIS faces the challenge of requiring numerous techniques to minimize variance. Conversely, if the \(S\) vertex is purely specular, it leads to bias in SMIS. This is observed in the \emph{Projector} scene shown in our experiments.

Another case is when different lights do not share the same visibility, as shown in \autoref{fig:smis-proxy-1} (b). In scenes with multiple light sources possessing complex visibilities, SMIS tends to exhibit obvious bias. This corresponds to SMIS's first limitation and will be further demonstrated in the scene \emph{Hallway}. While MMIS can eliminate bias by combining SMIS with unbiased sampling techniques like unidirectional path tracing, the inherent high variance of SMIS makes the overall estimator inefficient.

The analysis above leads us to prefer reciprocal estimation over SMIS/MMIS for our method, with experimental findings detailed later. We also highlight that our reciprocal estimator can be integrated into the CMIS framework, serving as a potent SMIS alternative. The CMIS estimator (\autoref{equ:balance CMIS definition}) is effectively computed using reciprocal estimation, which shows superior performance in certain scenarios. This is exemplified in our application to photon plane sampling \cite{photon_surface}, a method formerly using SMIS \cite{CMIS}. Detailed results in Sec. \ref{sec:photon plane} highlight our method's advantages.

\section{Proxy Tracing for BDPT}
With the proxy sampling method and reciprocal estimation discussed above, we therefore can improve the performance of probabilistic BDPT with specular connection by modifying the problematic vertices in the path sampled by specular connection. If the eye sub-path is on the specular surface, we can always continue the tracing of the eye sub-path and use the MIS method to get a robust sampling. Therefore, We focus on the case where the light sub-path is on the specular surface. 

\subsection{Proxy Construction}
When a light sub-path $\bar{y}$ terminates at a specular vertex $y_{s-1}$ and needs to be connected with an eye sub-path $\bar{z}$, we keep the eye sub-path $\bar{z}$ and the specular vertex $y_{s-1}$ unchanged in the proxy sampling. Then, we retrace the light sub-path $\bar{y}$ based on the location of $z_{t-1}$ that can satisfy specular restriction.

Not all the vertices in the light sub-path $\bar{y}$ need to be retraced. To ensure a valid BSDF at vertex $y_{s-1}$, we only retrace $y_{s-2}$ by sampling the BSDF $B(z_{t-1}\rightarrow y_{s-1}\rightarrow y_{s-2})$. However, for a light sub-path that ends with multiple consecutive specular vertices, only retracing $y_{s-2}$ is insufficient to satisfy the multiple specular constraints in the path. Therefore, if there are $u$ consecutive specular vertices $y_{s-1}y_{s-2}...y_{s-u}$ at the end of a light sub-path, we will retrace $u$ vertices $\bar{g} = y_{s-2}y_{s-3}...y_{s-u-1}$ to complete the proxy sampling. 

The specular or diffuse attributes of each vertex in the sequence in a proxy path should be consistent with those of the original sequence in the original path that has been substituted. Otherwise, there could be multiple possible proxy 
paths $\bar{g}$ for a given full path $\bar{x}$. For instance, an $L_{DD}DSDE$ path $y_{0}y_{1}y_{2}z_{1}z_{0}$ can be sampled from either $\bar{g} = y_{1}$ or $\bar{g} = y_{1}y_{2}$ if the original $L_{DD}SSDE$ path is retraced. Dealing with multiple alternative paths $\bar{g}$ is difficult and impractical. So, we stop the retracing and discard the connection when the retraced alternate path $\bar{g}$ has a specular/diffuse mismatch with the original. This ensures that the alternate path $\bar{g}$ as the proxy for a connection can be uniquely determined. 

 \begin{figure}[t]
	\centering
	\includegraphics[trim={0cm, 0cm, 0cm, 0cm},clip,width=0.95\linewidth]{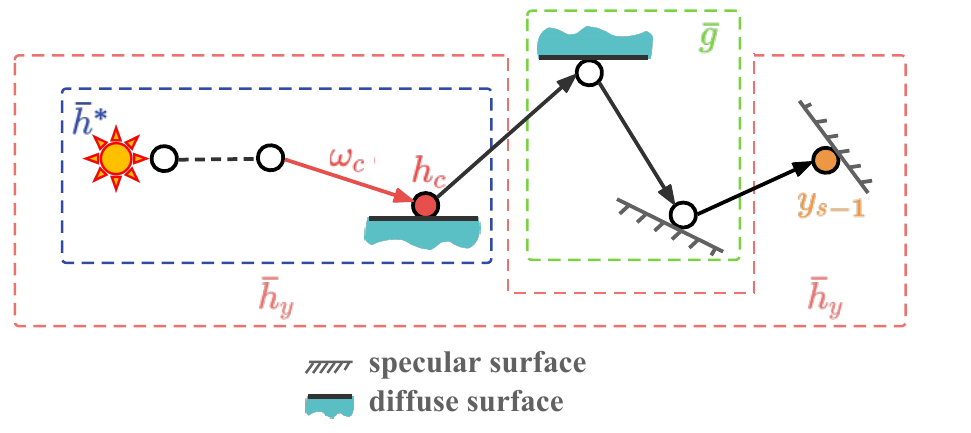}
	\caption{
A light sub-path can be segmented into several components: the path to be dropped out $\bar{g}$, the residual path $\bar{h}^*$, and the terminal specular vertex $y_{s-1}$ located on a specular surface. The control vertex $h_c$ represents the terminal of $\bar{h}^*$ and the control direction $\omega_c$ is the incident direction at $h_c$. Reciprocal estimation is determined by these elements: the control vertex $h_c$, control direction $\omega_c$, specular vertex $y_{s-1}$, and $u$, which is the vertex count of $\bar{g}$.
}
	\label{fig:lightpath-part}
\end{figure}

\subsection{PDF Evaluation}

We discuss the PDF for the incomplete sub-path sampling $p(\bar{h})$. $\bar{h}$ consists of the complete eye sub-path $\bar{z}$ and the incomplete light sub-path $\bar{h}_{y}$. Eye sub-path tracing is irrelevant to the tracing of the light sub-path. Therefore $p(\bar{h})$ is the product of $p(\bar{h}_{y})$ and $p(\bar{z})$ as:
\begin{equation}
    p(\bar{h}) = p(\bar{h}_{y})p(\bar{z}).
\end{equation}
Since $p(\bar{z})$ can be accurately computed, we only need to make a reciprocal estimation of the incomplete light sub-path $1/p(\bar{h}_{y})$. We further analyze the tracing of incomplete light sub-path $\bar{h}_{y}$. A light sub-path $\bar{y} = R(\bar{h}_{y},\bar{g})$ can be divided into several parts, as shown in \autoref{fig:lightpath-part}. They are the alternate path $\bar{g}$, the precedent path before reaching the alternate path $\bar{h}^* = y_{0}y_{1}...y_{s-u-2}$, and the last specular vertex $y_{s-1}$. According to \autoref{equ:tracingPdf}, the probability density of a vertex $y_{i}$ being traced is determined by its precedent two vertices. Therefore, the probability density of tracing the light sub-path $\bar{y}$ can be 
\begin{equation}
    p(\bar{y}) = p[R(\bar{h}_{y},\bar{g})] = p(\bar{h}^*)p(y_{s-1}|\bar{g}, h_c)p(\bar{g}|h_c, \omega_c),
\end{equation}
where $h_c = y_{s-u-2}$ is the last vertex of $\bar{h}^*$,$\omega_c = y_{s-u-3}\rightarrow y_{s-u-2}$ is the incident direction of $h_c$. 
Under extreme conditions, $h_c$ is set to null when $y_{0}\in \bar{g}$, $\omega_c$ is set to null when $y_{0}\in \bar{g}$ or $y_{1}\in \bar{g}$.
$h_c$ affects the sampling of $y_{s-1}$ only when the vertex count of $\bar{g}$ equals one. There is no vertex before $\bar{h}^*$, so the tracing of $\bar{h}^*$ is irrelevant to the alternate path $\bar{g}$. Only $y_{s-u-2}$ and $y_{s-u-3}$ can make sense to the tracing of $\bar{g}$ and $y_{s-1}$. Therefore, we call $h_c$ and $\omega_c$ as the control vertex and control direction.

Then, according to \autoref{equ:incomplete_integral}, the PDF for the incomplete light sub-path $\bar{h}_{y}$ is as
\begin{equation}
    \label{equ:reciprocalFinal}
\begin{aligned}
    p(\bar{h}_{y})&=\int_{A^u}p[R(\bar{h}_y,\bar{g})]d\mu(\bar{g})\\&=p(\bar{h}^*)\int_{A^u}p(\bar{g}|h_c,\omega_c)p(y_{s-1}|\bar{g},h_c)d\mu(\bar{g})\\&=p(\bar{h}^*)\mathcal{P}(u,h_c,y_{s-1},\omega_c),
\end{aligned}
\end{equation}
where $\mathcal{P}(u,h_c,y_{s-1},\omega_c) = \int_{A^u}p(\bar{g}|h_c,\omega_c)p(y_{s-1}|\bar{g},h_c)d\mu(\bar{g})$ is the exact probability density that involves PDF integral and requires reciprocal estimation. 

Note that a reciprocal estimation is determined by the four variables included in \autoref{equ:reciprocalFinal}. They are the glossy count $u$, control vertex $h_c$, control direction $\omega_c$, and the specular vertex $y_{s-1}$. These variables can be determined during the light sub-path tracing and are completely irrelevant to the eye sub-path. Therefore, the reciprocal estimation can be done in the light sub-path tracing instead of the run-time connection. This allows us to employ light sub-path reuse \cite{LVCBPT} when tracing incomplete light sub-path $\bar{h}_{y}$. Specifically, we can trace and estimate the reciprocal for a small number of incomplete light sub-paths $\bar{h}_{y}$ and connect those incomplete sub-paths with eye sub-paths from all the pixels. By doing so, we can save a significant amount of cost associated with reciprocal estimation.

\subsection{Supporting Distribution Construction}
Our reciprocal estimation requires a supporting distribution $q$ to sample the proxy path $\bar{g}$ in the estimation of $\frac{1}{\mathcal{P}(u,h_c,y_{s-1},\omega_c)}$. We use three sampling strategies for the proxy path $\bar{g}$ sampling and apply these strategies to the supporting distribution $q(\bar{g})$.

\subsubsection{Light Source Sampling}
If the only vertex that requires retracing is on the light source, i.e., when the alternate path $\bar{g}=y_{0}$, we can simply sample the light source surface to obtain $y_{0}$. 
However, this method is disabled when $u>1$, as in the case where $u=2$ and alternate path $\bar{g}=y_{0}y_{1}$, with $y_{1}$ being sampled by tracing from $y_{0}$. To satisfy the constraint on $y_{1}$ (which has to be on a specular surface), we need to trace $y_{2}$ from $y_0 \rightarrow y_{1}$, but $y_2$ is the specular vertex and fixed, making it impossible. As a result, $p(y_{0}\rightarrow y_{1} \rightarrow y_{2})$ would be close to zero and the alternate path $\bar{g}$ would be ineffective for rendering.

\subsubsection{Tracing from the Control Vertex}
If $u=1$ and $y_{0}$ are not included in the alternate path $\bar{g}$, we trace the only alternate vertex $y_{s-2}$ from the control vertex $h_c$. If the control vertex is on the light source, i.e., $h_c = y_{0}$, we sample the cosine hemisphere space at $h_c$ to determine the tracing direction; otherwise, we sample the BSDF at $h_c$ using $\omega_c$ as the incident direction to determine the tracing direction. This method is also disabled when $u>1$ for the same reason as light source sampling.

\subsubsection{Tracing from the Specular Vertex $y_{s-1}$}
This method samples the alternate path $\bar{g}$ by tracing from the specular vertex $y_{s-1}$. The initial direction is sampled from the cosine hemisphere space of $y_{s-1}$. For each intermediate vertex $y_{s - i}$, continue the tracing for $y_{s-i-1}$ based on the BSDF at $y_{s-i}$ and direction $y_{s-i + 1}\rightarrow y_{s-i}$ until all $u$ vertices are traced. 

The final supporting distribution is the uniform mixture of the available sampling strategies. That is, if $u=1$ and $\bar{g}=y_{0}$, we use light source sampling and tracing from $y_{s-1}$ to sample the alternate path $\bar{g}$; otherwise, if $u=1$ we sample the alternate path $\bar{g}$ from the control vertex $h_c$ and the specular vertex $y_{s-1}$; for $u>1$, the only available strategy is to sample the alternate path from the specular vertex $y_{s-1}$.

\subsection{MIS Weighting Function for Reciprocal Estimation}
\label{sec:MIS_weight}
While our method provides an efficient strategy for sampling specular paths, it may not be efficient in all cases. For instance, when the end of eye sub-path $z_{t-1}$ is very close to the specular vertex $y_{s-1}$, a unidirectional path tracing that proceeds to trace the eye sub-path $z_{t-1}\rightarrow y_{s-1}$ may be more effective. Thus, it is necessary for our method to combine with the existing methods by multiple importance sampling. 

A well-known MIS weighting function is the balance heuristic \cite{BDPT_mis} that assigns MIS weight proportional to the PDF of each strategy. This heuristic can minimize the upper bound of variance of the combined estimator. However, the PDF-proportional balance heuristic is derivated when the PDF for each strategy can be uniquely determined. In the reciprocal estimation, our estimation of $\frac{1}{p(\bar{h})}$ is a random variable that may introduce extra variance to the combined estimator. Therefore, we need to derive the MIS weight for reciprocal estimation as follows.

Adopting the way similar to \citet{BDPT_mis}, our objective is to minimize $E(I^2)$, the upper bound of the combined estimator $I$ is as
\begin{equation}
    \label{equ:MIS_0}
    E(I^2) = \sum_t \int_{\Omega} w^{2}_{t}(\bar{x})f^{2}(\bar{x})E[\frac{1}{\tilde{p}^2_t(\bar{x})}]p_t(\bar{x})d\mu(\bar{x}),
\end{equation}
where $t$ refers to the identity of sampling strategy, $w_t$ is the corresponding MIS weighting function for strategy $t$, and $\frac{1}{\tilde{p}(\bar{x})}$ represents an unbiased estimation for $\frac{1}{p(\bar{x})}$. 

The optimal MIS weighting function for $w_t$ in \autoref{equ:MIS_0} is as:
\begin{equation}
    \label{equ:MIS_1}
    w_{t}(\bar{x}) = \frac{1/(E[\frac{1}{\tilde{p}^2_t(\bar{x})}]p_t(\bar{x}))}{\sum_{i} 1/(E[\frac{1}{\tilde{p}^2_i(\bar{x})}]p_i(\bar{x}))}.
\end{equation}
If $p_t(\bar{x})$ can be precisely computed without any variance, then $E[\frac{1}{\tilde{p}^2_t(\bar{x})}] = \frac{1}{p^2_t(\bar{x})}$ and \autoref{equ:MIS_1} reduces to the PDF-proportional balance heuristic. For the proxy sampling, its performance is decreased by the variance of reciprocal estimation, resulting in less MIS weight.

%We obtain the estimation of $E[\frac{1}{\tilde{p}^2_t(\bar{x})}]$ and $p_t(\bar{x})$ in reciprocal estimation according to the subspace statistic data discussed in \autoref{sec:subspaceStatistic}. 

\subsection{Subspace for Statistic}
\label{sec:subspaceStatistic}
According to Sec. \ref{sec:reciprocalEstimation}, the reciprocal estimation requires the upper bound $B$ for optimization. In Sec. \ref{sec:MIS_weight}, $E[\frac{1}{\tilde{p}^2_t(\bar{x})}]$ and $p_{t}(\bar{x})$ are required to compute the MIS weight. However, those statistical data are unavailable in the practical estimation. We need an approximation for the statistical data.  

In \autoref{equ:reciprocalFinal}, reciprocal estimation can be identified by four variables: the specular vertex $y_{s-1}$, the control vertex $h_c$, the control direction $\omega_c$, and the glossy count $u$. We adopt the same mechanism of subspace proposed by Su et al.~\citet{SPCBPT} and trace a small number of $L(S|D)^*SDS^*E$ paths by PT at the beginning of rendering. The control and specular vertices from these paths are used to build control subspace and specular subspace, respectively, to classify $h_c$, $\omega_c$, and $y_{s-1}$. A control subspace maps $(h_c,\omega_c)$ to control subspace label $C$; a specular subspace maps $y_{s-1}$ to a specular subspace label $S$. This allow us to approximate $\mathcal{P}(u,h_c,y_{s-1},\omega_c)$ as 
\begin{equation}
    \label{equ:subspaceApproximate}
    \mathcal{P}(u,h_c,y_{s-1},\omega_c)\approx \mathcal{P}(u,C,S).
\end{equation}
We do not use $\mathcal{P}(u,C,S)$ directly to estimate the reciprocal. Instead, we compute and store the statistical data for reciprocal estimation based on $(u,S,C)$ and these statistical data are used to support rendering tasks such as selecting upper bound $B$ and computing MIS weight. 

\subsection{Probabilistic Light Selection in Subspace} 
Our algorithm enables sampling of a specular path by connection strategy, which can further improve rendering efficiency when combined with existing probabilistic connection algorithms. However, some adaptations should be made to these connection algorithms for proper integration.

Most of the connection algorithms rely on the local contribution $f_{y}(\bar{y})$ of light sub-path to construct their resampling distribution. However, retracing of the proxy vertices may significantly alter the light sub-path, causing the $f_{y}(\bar{y})$ of the original sub-path to differ greatly from the new proxy sub-path. Therefore, connection algorithms should not use the $f_{y}(\bar{y})$ of the original sub-path in their contribution construction.

In our implementation, we employ the Subspace-based Probabilistic Connection for BPT (SPCBPT) \cite{SPCBPT} to select the appropriate light sub-path for proxy sampling. We utilize the specular subspace discussed in section ~\ref{sec:subspaceStatistic} to cache the incomplete light sub-paths and resample them based on their respective specular subspace. 

A subspace sampling matrix $\Gamma[T,S]$ is required to determine the PMF for an eye subspace $T$ to sample the specular subspace $S$ in the first stage of subspace sampling. We construct the subspace sampling matrix based on the expected contribution of the proxy path. Specifically, $\Gamma[T,S]$ is proportional to the expected contribution for an incomplete sub-path $\bar{h}_{y} \in S$ and an eye sub-path $\bar{z}\in T$. We learn the $\Gamma$ matrix on-the-fly during rendering. In the second sampling stage to sample the sub-path from the subspace, the sub-path is simply sampled uniformly in the specular subspace $S$ to obtain the incomplete sub-path for the subsequent proxy sampling.
 
\section{Experiments and Results}

We validate our approach in various scenarios and make comparisons with unidirectional path tracing (PT) with MIS combination of next event estimation (NEE) as the baseline, the state-of-the-art probabilistic connections method SPCBPT~\cite{SPCBPT}, an SPCBPT with Light Tracing (LT) enabled, and an LVCBPT~\cite{LVCBPT} with LT enabled, which is a GPU accelerated version of BDPT that serves as a BDPT baseline. Our method is implemented based on SPCBPT and uses SPCBPT to handle the non-specular paths and focuses on the sampling of the specular involved path. Mean absolute percentage error (MAPE) is used as the metric for comparison. We also provide the comparison and combination with path guiding of \citet{PG_SD}, and the effect of subspace sampling.
In Sec. \ref{sec:compare with SMIS in experiment}, we provide additional results comparing SMIS/MMIS sampling with our reciprocal estimation technique in proxy tracing. Furthermore, in Sec. \ref{sec:photon plane}, we showcase the application of reciprocal estimation as a substitute for SMIS in photon plane volume rendering, achieving superior performance in scenarios where SMIS previously excelled.

\begin{figure*}
	\centering
	\includegraphics[trim={4cm 3cm 3cm 8cm},clip, width=0.98\linewidth]{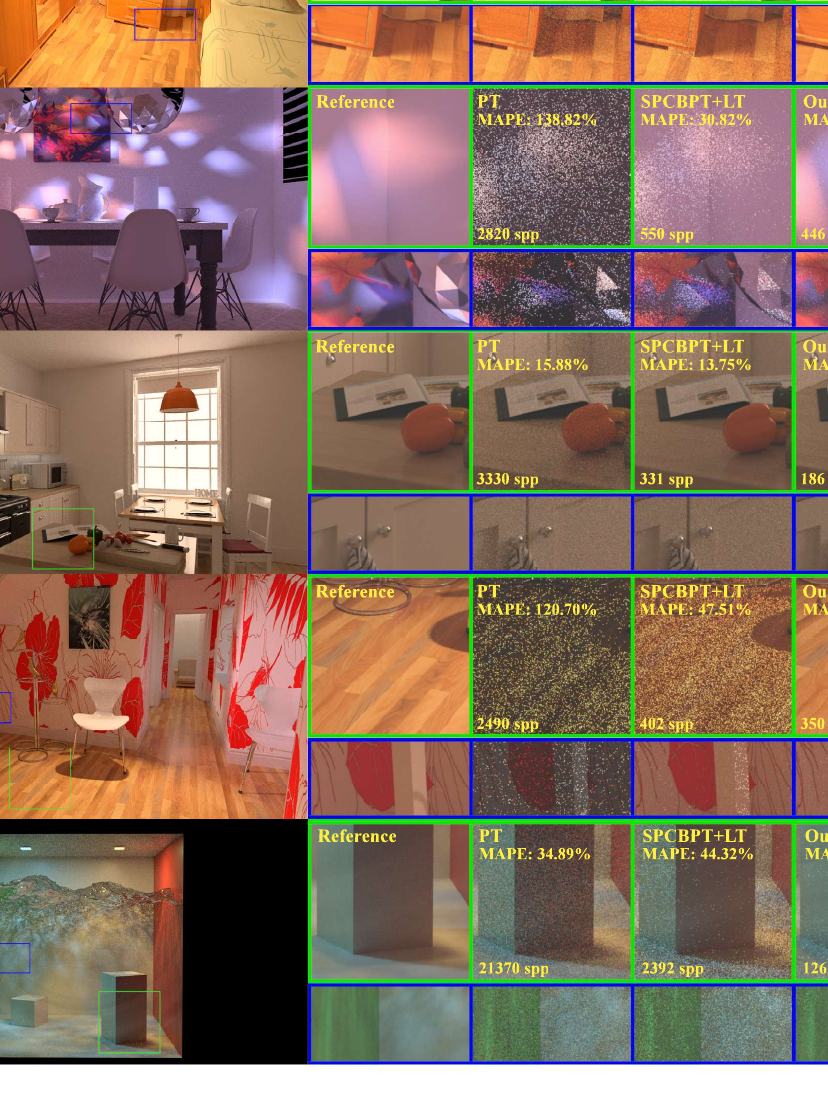}
	\caption{ Equal-time comparison: PT with NEE integration, SPCBPT~\cite{SPCBPT}, SPCBPT with LT enabled, and our approach. Rendering time cost is $180s$ for \emph{Water} and $60s$ for other scenes. The leftmost side shows the full frame synthesized by our approach. The number of Iterations (SPP, samples per pixel) and MAPE are shown in the zoom-in region. Our approach shows superior performance in terms of MAPE and produces less noisy results than other approaches.
    }
	\label{fig:res_all}
\end{figure*}

\subsection{Experimental Setting}

\begin{figure*}
% \centering
\includegraphics[trim={10cm 4cm 9cm 4cm},clip,width=0.9\linewidth]{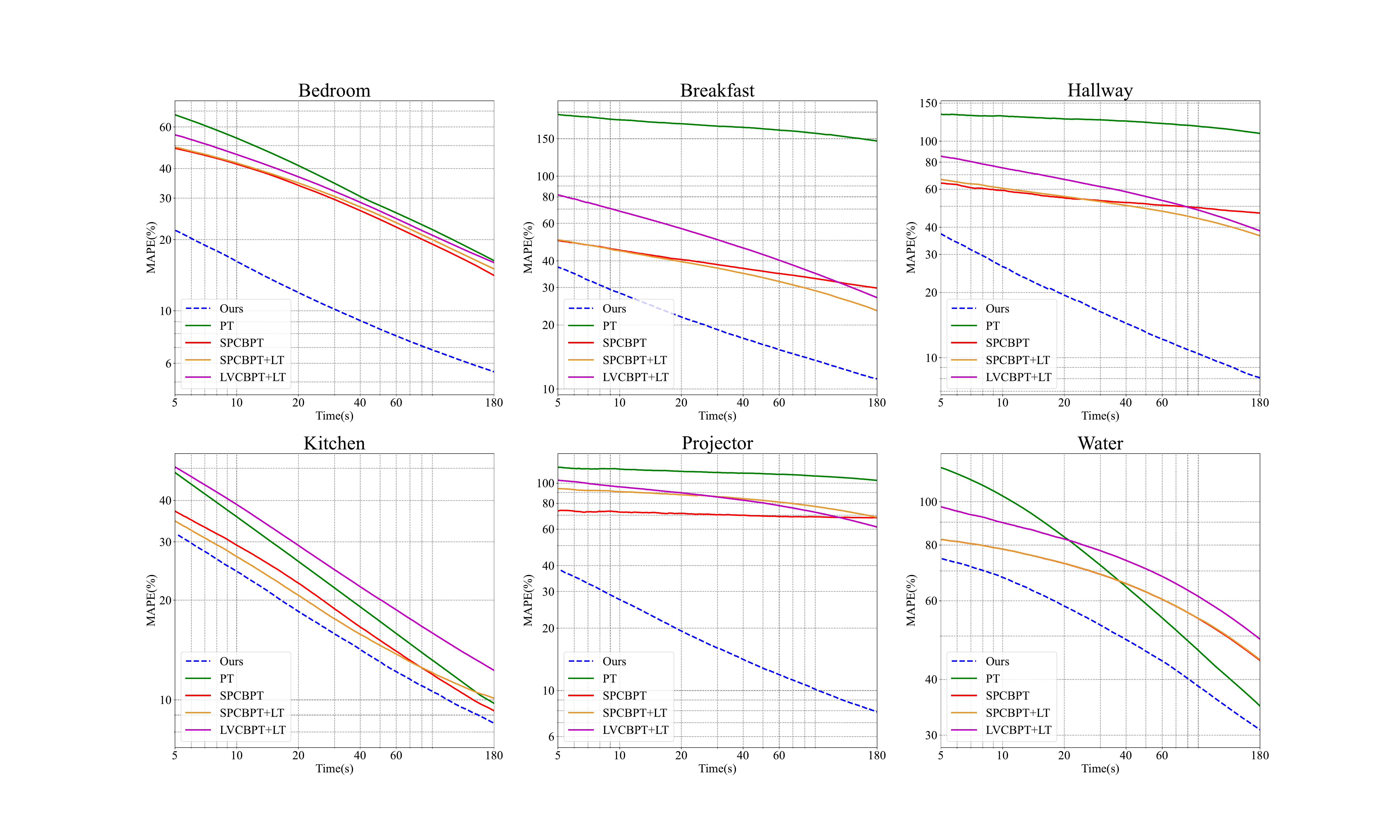}
\caption{Convergence over time (3 min) in terms of MAPE for PT, SPCBPT~\cite{SPCBPT}, SPCBPT with LT, LVCBPT \cite{LVCBPT} with LT,} and ours. Our approach shows superior convergence performance in all tested scenarios. In \emph{Water}, the SPCBPT curve overlaps with the SPCBPT+LT curve because a specular and transparent surface seals the box, making LT have no contribution. 
\label{fig:loglog}
\end{figure*}

\begin{table*}[t]
\centering
\caption{
Time cost (measured in seconds) and iterations to reach a specified MAPE by different methods. Our approach shows the best performance consistently and outperforms others significantly in both time cost and iteration required. }
\begin{tabular}{ c|cc|cc|cc|cc|cc|cc }
\bottomrule
\multirow{3}{*}{\diagbox{}{Scene}} &\multicolumn{2}{c|}{Bedroom} & \multicolumn{2}{c|}{Breakfast} &\multicolumn{2}{c|}{Hallway} &\multicolumn{2}{c|}{Kitchen}
 &\multicolumn{2}{c|}{Projector} &\multicolumn{2}{c}{Water} \\
 
 &\multicolumn{2}{c|}{(20\%)} & \multicolumn{2}{c|}{	(30\%)} &\multicolumn{2}{c|}{(35\%)} &\multicolumn{2}{c|}{(10\%)}
 &\multicolumn{2}{c|}{(20\%)} &\multicolumn{2}{c}{(35\%)} 
\\
\cline{2-13}
&Iter. &Time &Iter. &Time &Iter. &Time  &Iter. &Time  &Iter. &Time  &Iter.  &Time  \\
\hline
PT & 6533 & 113.35
& >14324 & >200
& >7631 & >200 
& 9248 & 168.42
& >11071 & >200 
& 21225 & 178.37 \\

\hline
SPCBPT & 582 & 81.05 
& 1529 & 171.45 
& >1295 & >200
& 788 & 142.44
& >1357 & >200
& >2691 & >200 \\

\hline
SPCBPT+LT & 683 & 90.22
& 609 & 78.68
& >1311 & >200
& 1052 & 189.38
& >1412 & >200
& >2666 & >200 \\

\hline
LVCBPT+LT & 983 & 100.27
& 1739 & 135.54
& >1790 & >200
& >1358 & >200
& >1844 & >200
& >2783 & >200 \\

\hline
\textbf{Ours} & \textbf{29} & \textbf{6.23}
& \textbf{62} & \textbf{8.63}
& \textbf{27} & \textbf{5.73}
& \textbf{337} & \textbf{107.26}
& \textbf{70} & \textbf{18.93}
& \textbf{829} & \textbf{121.32} \\

\toprule
\end{tabular}
\label{tab:comparison}
\end{table*}
\subsubsection{Renderer Setting}
All the algorithms are implemented based on the OptiX architecture \cite{parker2010optix}, and run on an NVIDIA GeForce RTX 3090 GPU with an Intel Core i9-12900K CPU on a Windows system. All images are rendered in a resolution of $ 1920 \times 1000$. We build $300$ eye subspaces and $300$ light subspaces for SPCBPT. Compared to the default setting of $1000$ subspaces in SPCBPT, a smaller number of subspaces can avoid overfitting and works better for scenes with easy visibility, as discussed in \citet{SPCBPT}. We use PT to pretrace $1,000,000$ full paths to train the sampling distribution of SPCBPT. In each iteration, SPCBPT, LVCBPT and our method all trace $M=10,000$ light sub-paths for resampling. Our approach shares the same parameter and configuration with SPCBPT as recommended in~\citet{SPCBPT}, unless otherwise stated. 

In our proxy sampling method, we impose constraints on the path processing. We drop out the problematic vertices and obtain the incomplete light sub-path only if there is no control vertex or the control vertex is on the light source. This constraint limits the light sub-path type to $L_{DD}S^+$ and $L_{DD}DS^+$, as the contribution of $L_{DD}D^+S^+$ is negligible and not visible in the rendered image. Additionally, the glossy count $u$ is limited to $1\le u\le4$. Paths that do not satisfy these constraints are handled by SPCBPT. In each iteration, we perform reciprocal estimation on up to $400$ incomplete light sub-paths and discard the others to reduce overhead as much as possible. The incomplete sub-paths are divided into $100$ specular subspaces for probabilistic connections. We use only $10$ subspaces for the control vertex since the control vertex is limited to the light source. To reduce the variance of reciprocal estimation, we repeat the reciprocal estimation $5$ times for each incomplete light sub-path and use the average as the final estimation. We learn the subspace sampling matrix in the first $40$ iterations and then freeze the matrix for stable rendering performance. We only sample the specular light sub-path and perform proxy sampling when the eye sub-path travels to its first diffuse surface. Additionally, we divide the image into grids, with each grid containing $10 \times 10$ pixels. If the contribution of our target path exceeds $5\%$ in a grid, we sample the incomplete light sub-path and perform the proxy sampling in probability $100\%$; otherwise, the probability is set to $20\%$.  

We adopt the DISNEY principled BSDF \cite{disney2012bsdf} for the BSDF model of highly glossy material. The roughness of most of the specular material in our experiment is set to $0.01$. A material is identified as specular when the material is metallic or transparent with roughness smaller than $0.2$.

 \begin{figure}[t]
     \centering
     \includegraphics[trim={3cm 2cm 0.8cm 0.8cm},clip, width=0.99\linewidth]{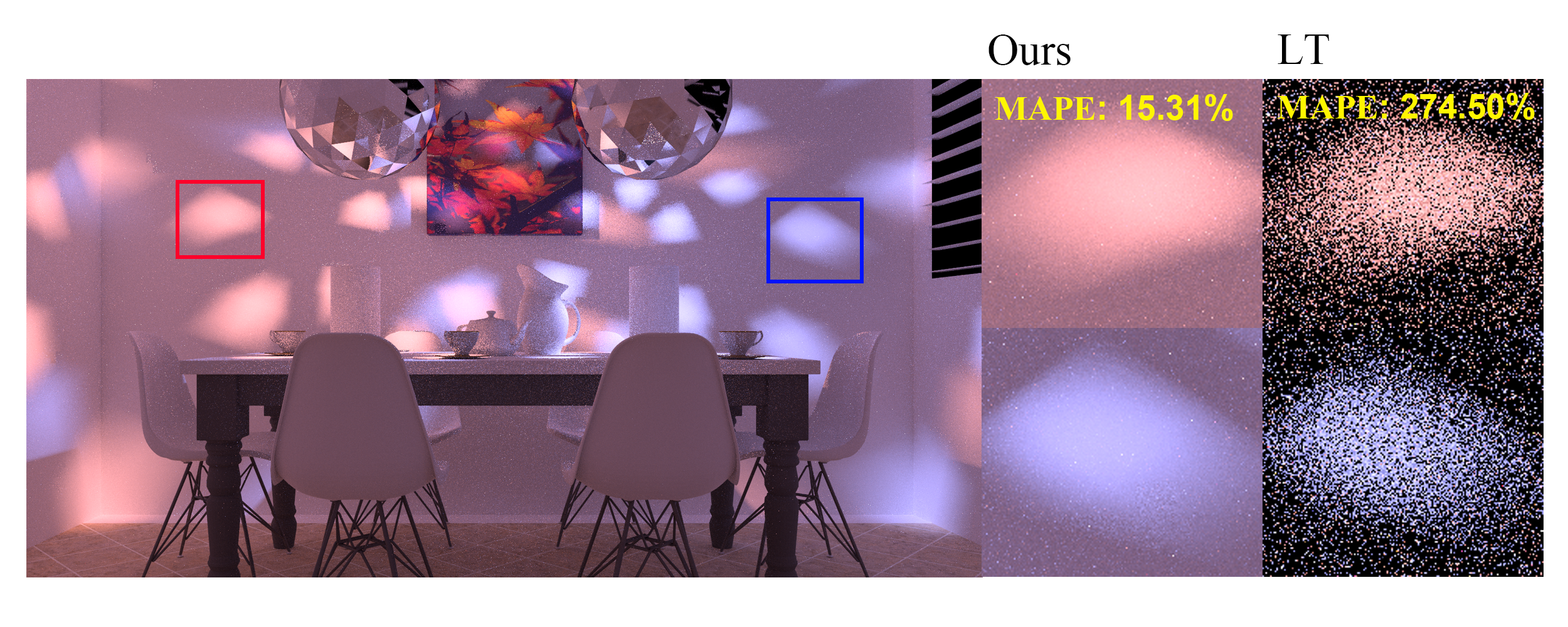}
     \caption{Comparison between Light Tracing (LT) and our approach in \emph{Breakfast}. Our approach delivers superior rendering results in 60$s$, whereas LT takes 180$s$. This showcases the efficiency of our method in efficiently sampling challenging paths and producing high-quality caustic patterns in significantly less time than LT. }
     \label{fig:LT-compare}
 \end{figure}

\subsubsection{Scene Settings}
Six benchmarks, including \emph{Bedroom}, \emph{Breakfast}, \emph{Projector}, \emph{Hallway}, \emph{Kitchen}, and \emph{Water}, are tested in our experiment. These benchmarks cover different illumination conditions and glossy surface settings.  

The \emph{Bedroom} scene is illuminated by a light source positioned above a mirror. Some of the light directly illuminates the scene, while the other light is reflected off the mirror and into the scene as our target path. The Mirror on the wall shows the effect of $L_{DD}SDSE$ path. In the \emph{Breakfast} scene, two glossy angular spheres are positioned above two cylindrical light sources on the table to create red and blue caustic patterns on the wall. The \emph{Projector} scene features a textured light source enclosed in a box with a convex lens, creating a projector effect. The convex lens refracts the textured light before projecting it onto the wall. \emph{Hallway} is mainly illuminated by two light sources enclosed in a glass lamp, requiring the light to travel through the refractive glass to enter the scene. In the \emph{Kitchen} scene, the light comes from the outside and is located behind a two-layer glass window. Finally, the \emph{Water} scene displays a classic caustic pattern created by small light sources and complex refractive surfaces.

The preprocessing times for SPCBPT to pre-trace (using PT) the full paths never exceed $5$ seconds, which depends on the efficiency of PT. On average, each pixel only needs to trace approximately $0.5$ path to obtain $1,000,000$ full paths for preprocessing. This time cost is negligible compared to the convergence time of PT. Therefore, the preprocessing time is not taken into consideration during our comparisons. 
\emph{Projector}, \emph{Hallway}, and \emph{Water} are rendered by PT in $10M$ iterations and the rest scenes are rendered by SPCBPT in $100K$ iterations. 

\begin{figure}
	\centering
	\includegraphics[trim={0.0cm 1.9cm 0.0cm 2cm},clip, width=0.98\linewidth]{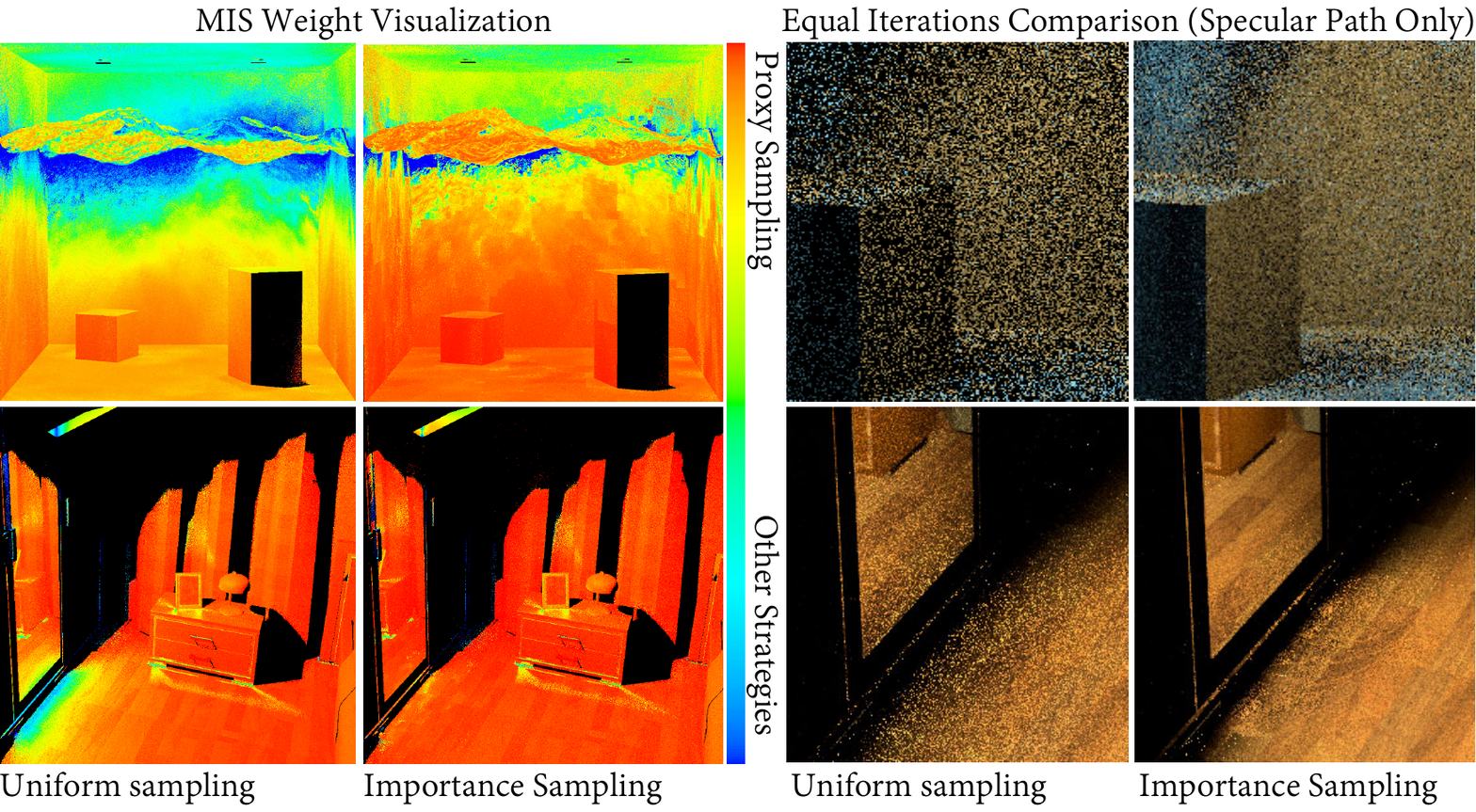}
	\caption{MIS weight visualization and equal iterations (32 iterations for \emph{Bedroom}, 128 iterations for \emph{Water}) comparison for our method with/without subspace-based importance sampling for the incomplete light sub-path. Overall MAPE for uniform sampling vs. importance sampling is 69.13\% vs. 54.98\% in \emph{Water} and 20.96\% vs. 19.74\% in \emph{Bedroom}.}
	\label{fig:subspaceCompare}
\end{figure}

\subsection{Performance Evaluation} 
We show the performance of different methods in~\autoref{fig:teaser} and~\autoref{fig:res_all}. We also highlight the zoom-in regions for a detailed comparison. When rendering difficult specular-involved paths, both PT and SPCBPT can only rely on the sampling of the unidirectional path tracing. However, within the given time budget, PT makes a much higher number of iterations than that of SPCBPT due to its very low overhead. Therefore, PT can sample this specular path more efficiently than SPCBPT. Conversely, SPCBPT can select the appropriate light sub-path in the connection and produce much better results in indirect illumination. The performance difference is evident in the highlight region of \emph{Breakfast} and \emph{Hallway} at \autoref{fig:res_all}. Our method focuses on sampling the specular-involved path and utilizes the strategy from SPCBPT to sample the residual path. In most cases, we can achieve better rendering performance on difficult paths like $L_{DD}(S|D)^*SDS^*E$ than PT and SPCBPT using the same time budget and fewer iterations. Our method takes extra computational overhead and the rendering speed for each iteration of our method is slower than SPCBPT at around $25\%\sim50\%$. The extra overhead comes from reciprocal estimation, retracing for the proxy vertices, and the MIS weight computation.

\autoref{fig:loglog} shows the convergence over time in terms of MAPE in all scenarios. Our approach outperforms others significantly in terms of convergence performance compared to other methods. In particular, in \emph{Breakfast}, \emph{Projector}, and \emph{Hallway}, neither PT nor SPCBPT is able to handle all the light paths well, resulting in slow convergence over time. In contrast, our method can effectively sample the $L_{DD}(S|D)^*SDS^*E$ path while allowing sufficient time for other advantageous sampling strategies from SPCBPT to contribute to the indirect illumination, ensuring stable convergence. While LT can handle paths with type $L_{DD}SDE$, like the reflected light on the wall in \emph{Breakfast}. However, our method achieves better performance than SPCBPT with LT, demonstrating its efficiency in sampling these paths. The \emph{Breakfast} scene is typically considered manageable with a pure LT algorithm. As the BDPT baseline, LVCBPT+LT generally shows poorer performance than SPCBPT+LT, as shown in \autoref{fig:loglog} and \autoref{tab:comparison}. In \emph{Projector}, the subspace-based probabilistic sampling will struggle with specular paths, and LVCBPT+LT exhibits a slight advantage due to faster iteration speeds. However, in other test scenes, results show that SPCBPT+LT outperforms LVCBPT+LT, which is consistent with the experimental results reported in \citet{SPCBPT}. Our method shows the best performance in all scenes.

To further evaluate our algorithm, we conducted an additional experiment comparing LT with our approach, as shown in \autoref{fig:LT-compare}. Remarkably, our method outperformed LT, delivering superior rendering results in just 60s, while LT required 180s but with much noise. This highlights the efficiency of our method in effectively sampling challenging paths and producing high-quality caustic patterns in significantly less time than LT. 

We also report the statistics of time cost and iterations
required to reach the specified MAPE in \autoref{tab:comparison}. Our approach demonstrates higher efficiency over SPCBPT, SPCBPT with LT, and PT, although the improvement varies across different scenes. For instance, in scenarios illuminated by large light sources like \emph{Kitchen}, tracing the direct illumination of specular light sub-path is relatively easier for PT. So, our improvement is less impressive. In contrast, in a challenging scene where the light source is very small, it becomes extremely difficult for unidirectional path tracing to find the light source. Our approach is particularly advantageous.

\subsection{Sampling in Subspace}
In our approach, subspace-based probabilistic sampling is used to select the incomplete light sub-path for proxy sampling. We learn the effect of subspace sampling through equal-iteration comparison and visualizing the MIS weight function for proxy sampling. The comparison for our method with/without subspace-based probabilistic sampling is shown in \autoref{fig:subspaceCompare}. We focus on the sampling of the specular-involved $L_{DD}(S|D)^*SDS^*E$ path and disable the rendering for the other paths. MIS weight is assigned based on the quality of proxy sampling, and the red regions indicate the advantage area for our method. Generally, the efficiency of proxy sampling is highest when the diffuse surface is far from the specular surface; thus, the improvement of subspace-based probabilistic sampling is relatively small in this case (overall MAPE $20.96\% \rightarrow 19.74\%$ in \emph{Bedroom}). However, in a more complicated scene setting where the diffuse surface is close to the specular surface, the probabilistic connections method can help find the appropriate incomplete path for connection and provide efficient enhancement for the specular path rendering (overall MAPE $69.13\% \rightarrow 54.98\%$ in \emph{Water}), thus, extending the capability of proxy sampling.

\begin{figure}[t]
	\centering
	\includegraphics[trim={4cm 0cm 4cm 1cm},clip, width=0.99\linewidth]{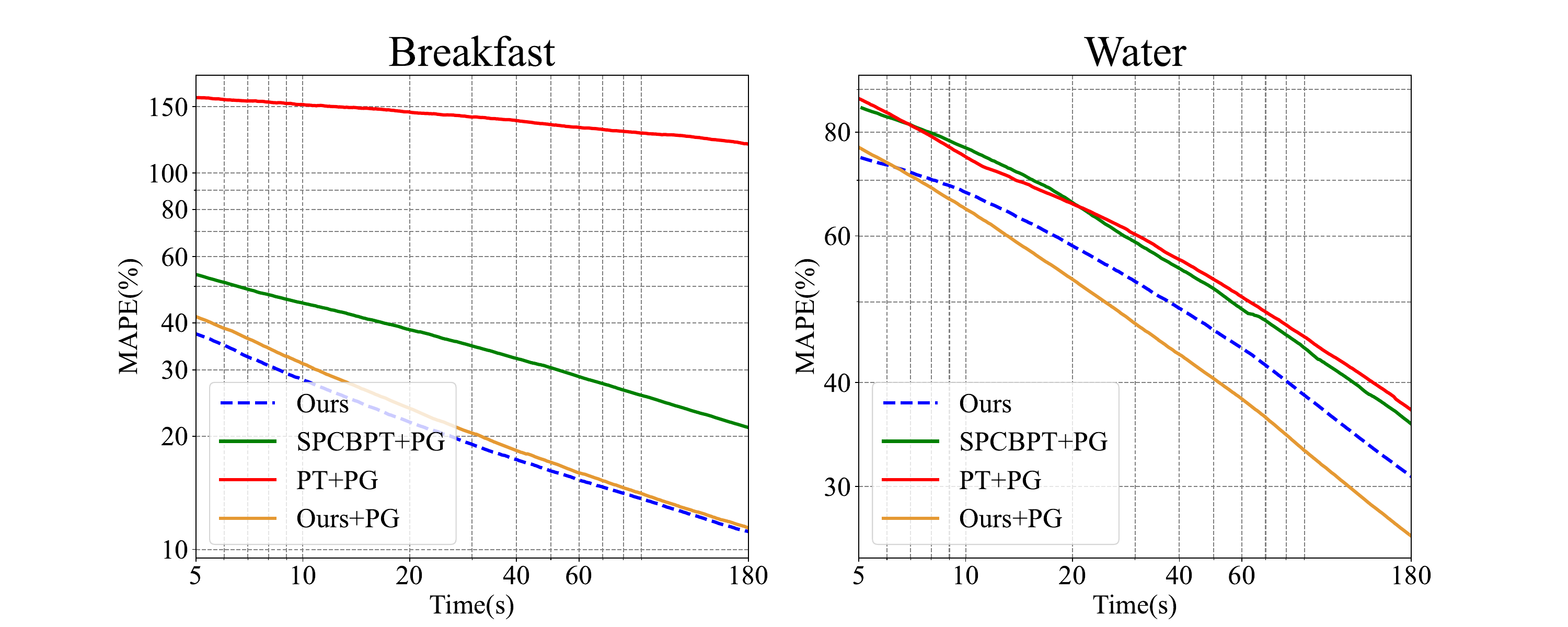}
 	 
	\caption{Convergence over time in terms of MAPE. We compare the performance of PT+Path Guiding (PG), SPCBPT+PG, our approach+PG, and our approach solely. $60s$ is used for \emph{Breakfast} and $180s$ for \emph{Water}. Path guiding is pre-trained for $30s$, and the pretraining time cost is not included. Our approach outperforms SPCBPT+PG significantly in \emph{Breakfast}, and achieves speedup when combined with PG in \emph{Water} scene.  }
	\label{fig:PGCompare_loglog}
\end{figure}

\begin{figure}
	\centering
	\includegraphics[trim={0cm 0cm 0cm 0cm},clip, width=0.99\linewidth]{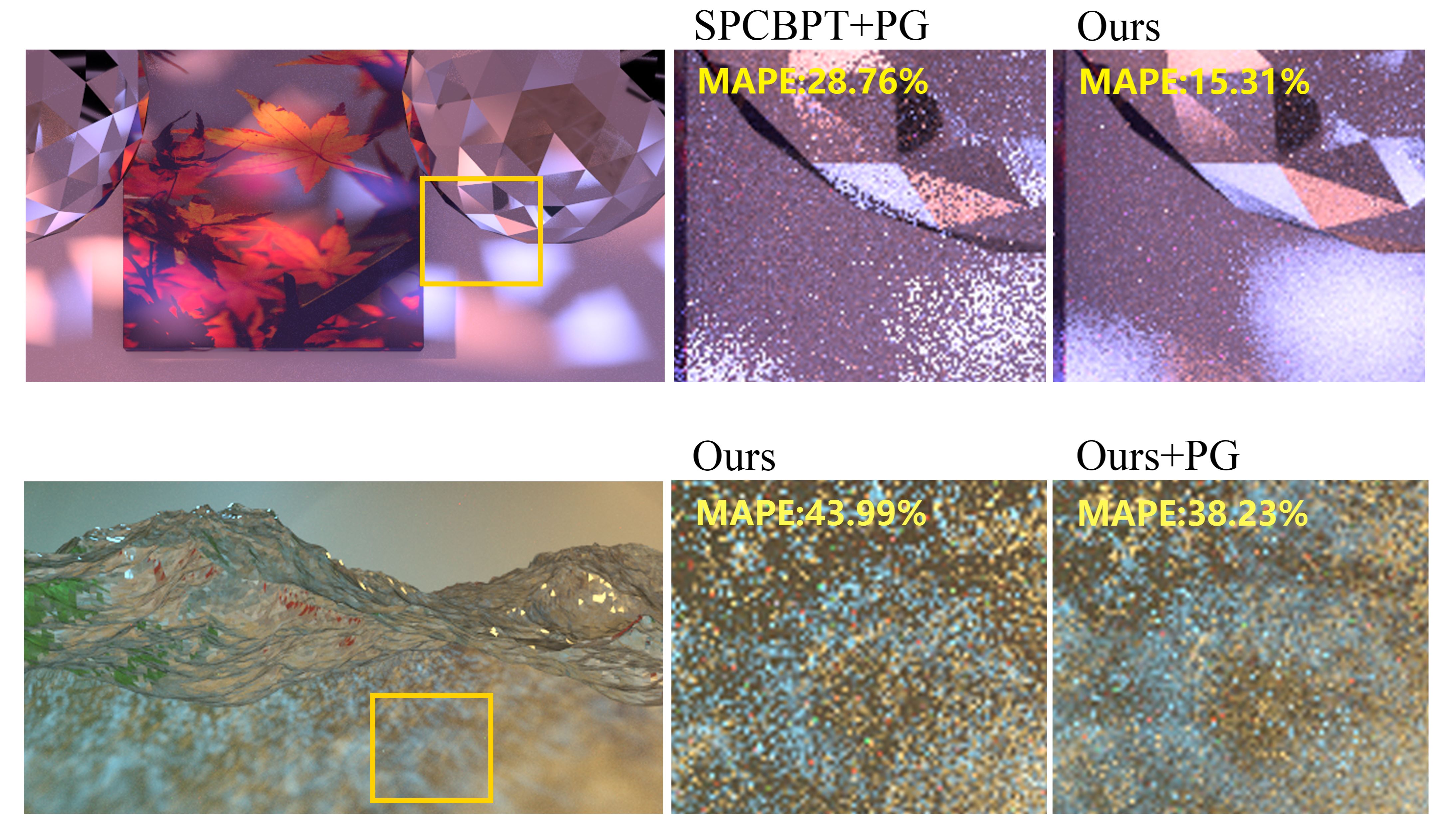}
 	 
	\caption{Rendering results obtained by equal time ($60s$ for \emph{Breakfast} and $180s$ for \emph{Water}) corresponds to \autoref{fig:PGCompare_loglog}. The upper row demonstrates the superiority of our approach over SPCBPT+PG; the lower row highlights the improvement of our algorithm through combination with PG in this \emph{Water} scene. }
	\label{fig:PGCompare}
\end{figure}

\subsection{Combination and Comparison with Path Guiding}
Our approach can be integrated well with the sophisticated path guiding technique. We integrate the path guiding method based on the SD-tree \citet{PG_SD} with our method and test its performance at \emph{Breakfast} and \emph{Water}. For each scene, we spend $30s$ to train the SD-tree for path guiding. We show the equal-time rendering with convergence plot in \autoref{fig:PGCompare_loglog}, and visual result in \autoref{fig:PGCompare}, time cost for pre-training not included. 

Since our approach can handle the specular-involved path in \emph{Breakfast} exceptionally well, it converges fast within just $60s$. Therefore, the speedup from path guiding is not significant and may even slightly slow down the performance of our approach, as shown in the \emph{Breakfast} scene. We also observed that our approach significantly outperforms SPCBPT+PG, possibly due to the small light source and the diffuse surface being far from the specular surface. In such case, a $30s$ training may not be sufficient for path guiding to help SPCBPT find the location of the light source quickly. 
In the scenario of \emph{Water}, the performance of our proxy sampling declines when the diffuse surface is very close to the specular surface (as indicated by the blue region in the MIS visualization of \autoref{fig:subspaceCompare}). Because in this case the specular vertex $y_{s-1}$ of the appropriate incomplete sub-path is usually very close to the eye sub-path and harder to select in the incomplete sub-path sampling, making our proxy sampling ineffective. In contrast, random direction sampling has a high probability of performing a good $D\rightarrow S$ bounce when the distance between diffuse surface $D$ and specular surface $S$ is very close. In such cases, combining our method with path guiding can gain benefits from both sides and improve performance, leading to even better results for this complicated setting. 

\begin{figure}[t]
	\centering
	\includegraphics[trim={4cm 1cm 4cm 1cm},clip, width=0.99\linewidth]{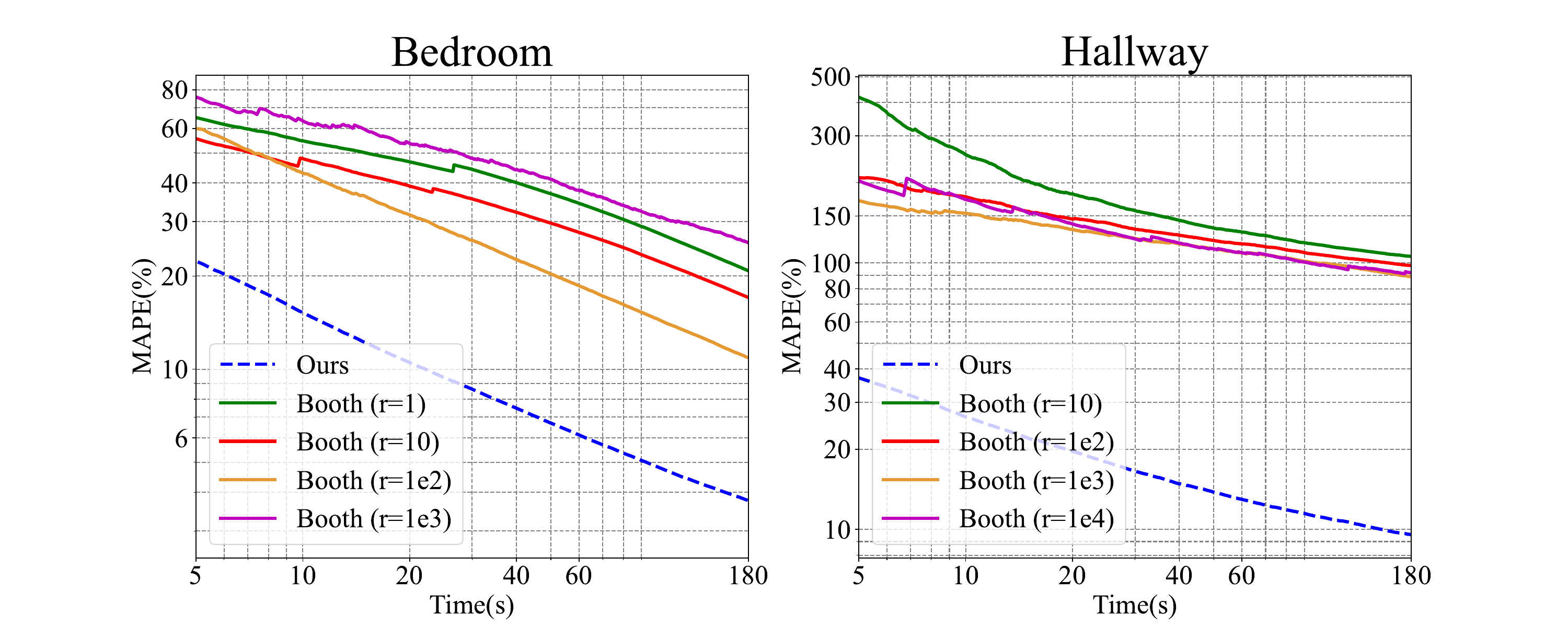}
\caption{Convergence over time in terms of MAPE. We perform an ablation study to assess the performance with and without our RR technique. The base algorithm is \citet{2007Booth} with parameter $r$ set at four different values. The results demonstrate the effectiveness and superiority of our RR technique.}
\label{fig:RR-ablation}
\end{figure}

\subsection{Ablation Study on RR Technique}
To assess the effectiveness of our RR technique, we conducted an ablation study, as shown in \autoref{fig:RR-ablation}. Due to the integration challenges with the method proposed by \citet{Taylor2015}, we opted for the framework described by \citet{2007Booth} as our baseline. Their RR strategy requires a manually set hyperparameter, $r$, and also necessitates an upper bound, $B$, which was the same in their experiments and ours. Our experimental findings reveal that the performance of the algorithm generally improves with an increase in $r$ up to a certain point, after which it diminishes. Specifically, the turning point occurs at an order of magnitude of 100 in the Bedroom scenario and at 1000 in the Hallway scenario, indicating that the optimal range for $r$ varies between different scenes. our results significantly outperform the baseline, irrespective of the $r$'s setting, suggesting that our RR method can effectively enhance the performance of reciprocal estimation.

\subsection{Comparison with SMIS/MMIS}
\label{sec:compare with SMIS in experiment}

\begin{figure*}[t]
    \centering
    \includegraphics[trim={7cm, 0cm, 7cm, 0cm},clip,width=\linewidth]{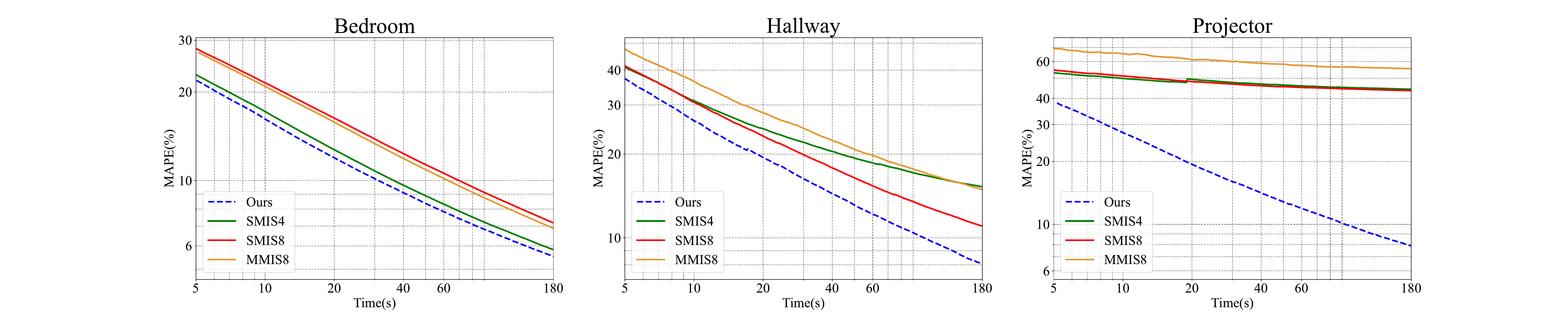}
    \caption{Convergence over time in terms of MAPE. We compare the performance of SMIS4, SMIS8, MMIS8, and our approach. $120s$ is used for each scene. Our approach outperforms SMIS/MMIS significantly in \emph{Hallway} and \emph{Projector}, and even in \emph{Bedroom} where SMIS/MMIS was considered to be advantageous.}
    \label{fig:compare-smis-loglog}
\end{figure*}

\begin{figure*}[htb]
    \centering

    \begin{minipage}{0.80\textwidth}
        \centering
        \includegraphics[trim={0.1cm, 0.1cm, 5.65cm, 0.1cm},clip,width=0.95\linewidth]{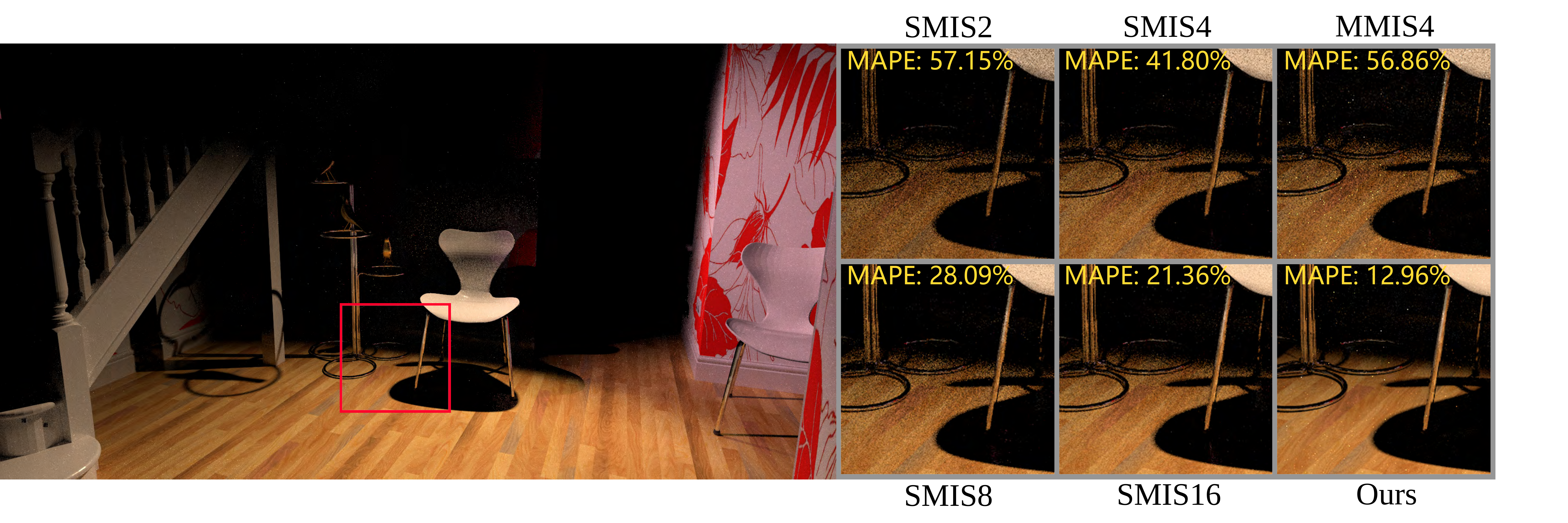}
    \end{minipage}
    \begin{minipage}{0.01\textwidth} {(a)} \end{minipage}
    
    \begin{minipage}{0.80\textwidth}
        \centering
    \includegraphics[trim={0.0cm, 0.2cm, 0.2cm, 0.2cm},clip,width=0.98\linewidth]{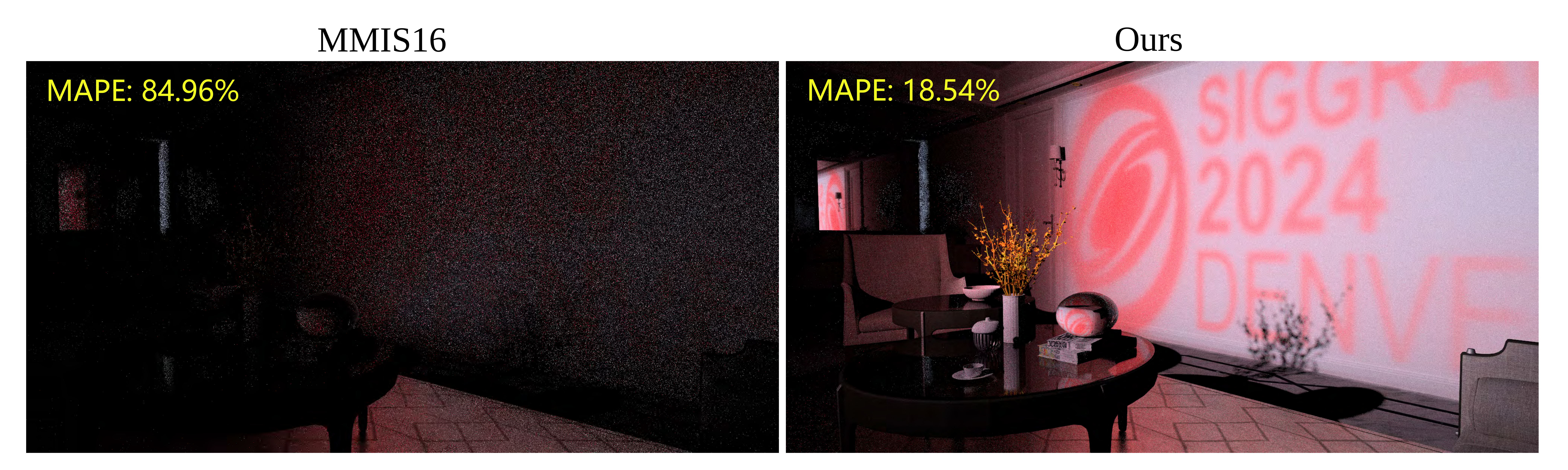}
    \end{minipage}
    \begin{minipage}{0.01\textwidth} {\ \newline (b)} \end{minipage}
        
    \caption{Comparison between SMIS/MMIS and our approach based on reciprocal estimation. We focus exclusively on rendering paths that are enhanced with proxy tracing, as the other components remain identical. (a) In \emph{Hallway}, SMIS$n$ displays obvious bias (darker) when $n$ is small. MMIS4 mitigates SMIS4's bias by combining unbiased techniques, the contribution of which appears as hot pixels. (b) In \emph{Projector}, SMIS/MMIS completely fails to capture the specular-involved difficult paths (see left), while our approach can handle this scene efficiently (see right).}
    \label{fig:hallway-smis-compare}
\end{figure*}

SMIS/MMIS can serve as a substitution for our reciprocal estimation in our proxy tracing approach. Therefore, we have integrated SMIS/MMIS into our proxy tracing framework and conducted a comparative analysis with our reciprocal estimation-based approach, using scenes such as \emph{Bedroom}, \emph{Hallway}, and \emph{Projector}. For this comparison, We only render the paths enhanced with proxy tracing, as the other parts remain consistent across methods. The $n$ in SMIS$n$ refers to the number of techniques combined in SMIS. We mainly compare with $n=4$ and $n=8$ cases here since a larger $n$ would be costly due to the $O(n^2)$ cost in the MIS process.  
\autoref{fig:compare-smis-loglog} illustrates the convergence over time in terms of MAPE in these scenes. Our reciprocal estimator outperforms SMIS/MMIS significantly. Reflecting on the limitations of SMIS/MMIS outlined in Sec. \ref{sec:comparison with CMIS}, it becomes evident that these constraints are noticeable when applied to scenes like \emph{Hallway} and \emph{Projector}.

In \emph{Hallway}, there are four light sources including an environment light. The scene is predominantly illuminated by two lights within glass lamps. It corresponds to the second failure case described in Sec. \ref{sec:comparison with CMIS}, where different lights have distinct visibilities. \autoref{fig:hallway-smis-compare} reveals the bias inherent in SMIS$n$. In SMIS2, SMIS4, and SMIS8, the illuminated areas appear noticeably darker compared to our unbiased approach. Increasing $n$ reduces this bias, making SMIS16 visually closer to our technique. However, note that there are only four light sources in this scene. In more complex scenes with hundreds or thousands of light sources, SMIS would require an impractically large $n$ to get good results. Additionally, we examined MMIS4 in \autoref{fig:hallway-smis-compare}. MMIS, which integrates SMIS with unbiased techniques like PT, mitigates the bias of SMIS. Specifically, the hot pixels in the output indicate other techniques' contribution. Nevertheless, as no technique effectively samples these paths, the overall efficiency remains sub-optimal.
In \emph{Projector}, the primary source of illumination is a projector that comprises a textured light source within a box and a convex lens. This setup aligns with the first failure case for SMIS, characterized by multiple specular vertices following the light vertex. In \autoref{fig:hallway-smis-compare}, we demonstrate the inability of SMIS/MMIS in this scenario.
Conversely, in \emph{Bedroom}, with its single light source and absence of consecutive specular surfaces, conditions are seemingly favorable for SMIS/MMIS. Despite this, our approach still significantly outperforms SMIS/MMIS, underscoring its high efficiency even in advantageous settings for the comparative methods.

\subsection{Reciprocal Estimation for CMIS Framework}
\label{sec:photon plane}

 \begin{figure}[t]    
	\centering
        \begin{minipage}{0.49\linewidth}
            \centering
            \includegraphics[width=0.80\linewidth]            {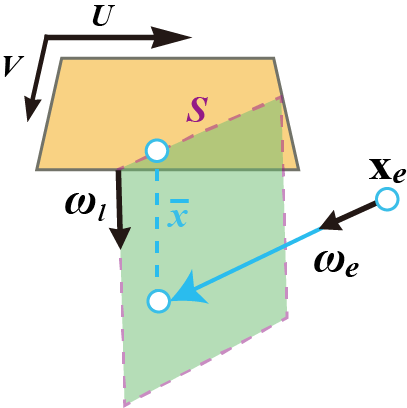}
            \centerline{(a)}
        \end{minipage}%
        \begin{minipage}{0.49\linewidth}
            \centering
            \includegraphics[width=0.80\linewidth]
            {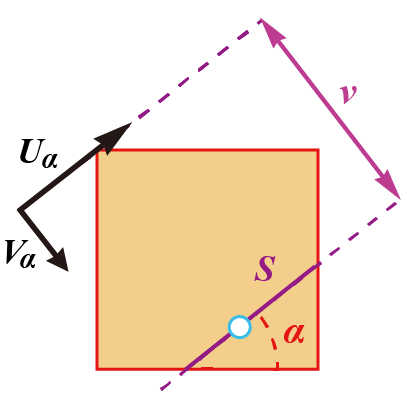}
            \centerline{(b)}
        \end{minipage} 
    \caption{(a) Sampling method of photon plane algorithm \cite{photon_surface}. Photon plane is generated by sampling a segment $s$ in the light source and an outgoing direction $\omega_l$. When an eye sub-path hits the photon plane, the camera, the intersection position, and the $\omega_l$-projection of the intersection position on the light source make the full path $\bar{x}$. (b) A segment $s$ can be sampled by first sampling a rotation angle $\alpha$ and then sampling the bias $v$.}
     \label{fig:photon plane alg}
\end{figure}

 \begin{figure}[t]    
    \centering
    \includegraphics[trim={1cm, 1cm, 1cm, 3cm},clip,width=0.99\linewidth]{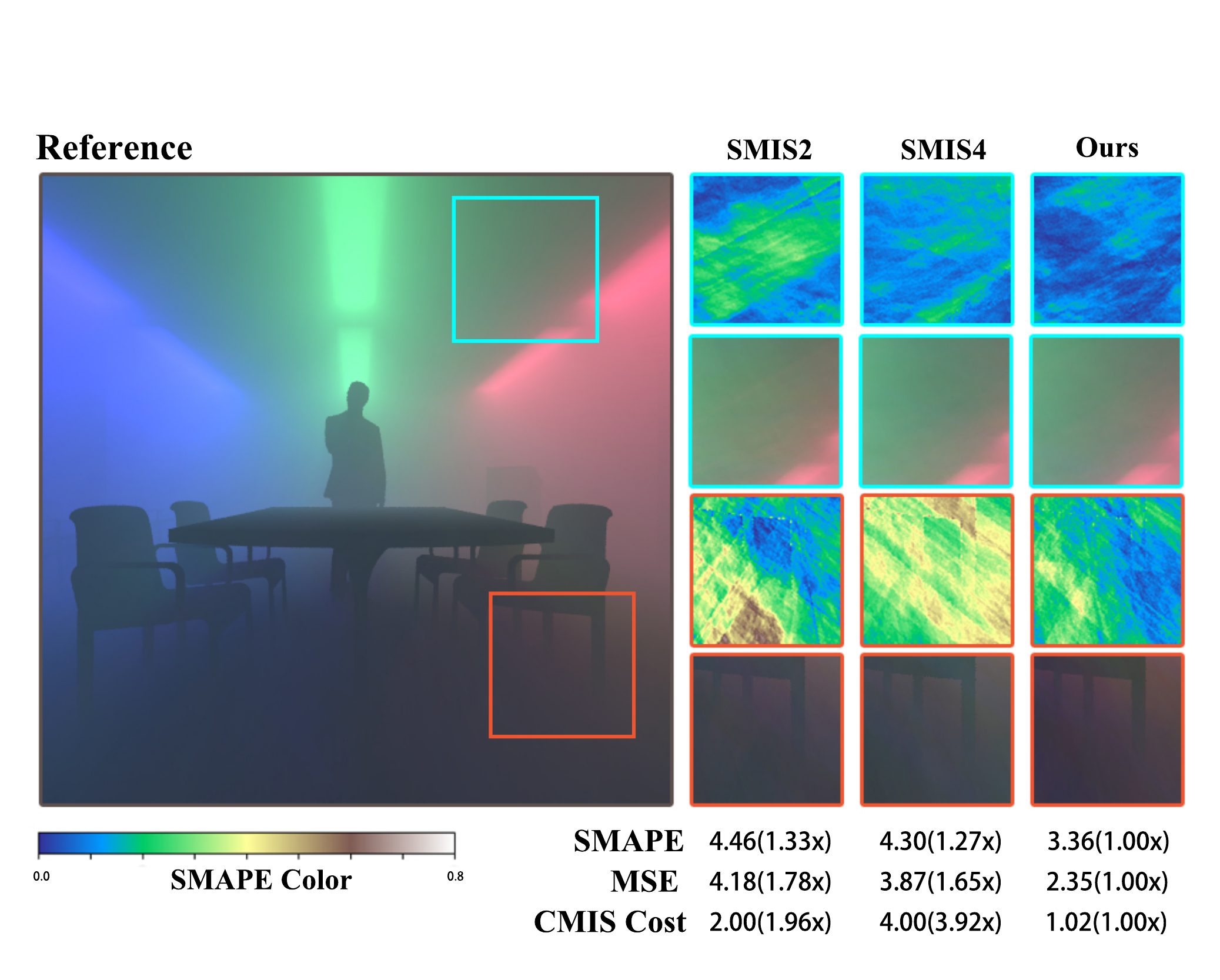}
    \caption{Equal-sample comparison between estimators for SMIS \cite{CMIS} and our method, specifically in volumetric scenarios rendered by photon plane algorithm \cite{photon_surface}. SMAPE is visualized by heat map and MSE below is magnified by $10^4$. \emph{CMIS Cost} refers to the average time cost of $p(\alpha, \bar{x})$ used for CMIS weight estimation.}
     \label{fig:photon plane}
\end{figure}

Our reciprocal estimation method can be applied to the CMIS framework since it can provide unbiased and efficient estimation for CMIS weight $\tilde{w}(t,x)$ in \autoref{equ:CMIS_weight_integral}. Here, we apply our method to the photon plane algorithm \cite{photon_surface}, which is an algorithm where CMIS excels. 

Photon plane is introduced by \citet{photon_surface} for single scattering volumetric rendering. The sampling method is shown in \autoref{fig:photon plane alg}(a). A photon plane is determined by a segment $s$ in the light source and an outgoing direction $\omega_l$. When an eye sub-path hits the photon plane, a light source vertex can be found by projecting the intersection vertex in direction $-\omega_l$, then a full path $\bar{x}$ can be sampled by connecting the camera, the intersection and the projected vertex on the light source.

Computing $p(\bar{x})$ is a challenge since the same $\bar{x}$ can be generated from different photon plane. \citet{CMIS} indicated that the sampling of the segment $s$ can be split into the sampling of rotation angle $\alpha$ and bias $v$, and $\alpha$ can be treated as the technique identifier in technique space $\mathcal{T}=[0,\pi]$ for CMIS framework, as shown in \autoref{fig:photon plane alg}(b).
The CMIS weight $\tilde{\omega}(\alpha, \bar{x}) = \frac{p(\alpha, \bar{x})}{\int_{\mathcal{T}}p(\alpha, x)d\alpha}$ involves the reciprocal of integral which is handled by SMIS in \citet{CMIS}. We replace SMIS by our reciprocal estimation method. The efficiency of reciprocal estimation depends on the supporting distribution for sampling $\alpha$. The optimal supporting distribution $q^*(\alpha|\bar{x})$ depends on $\omega_e$ as well as the position of the vertex on the light source. Therefore, we cache the piece-wise approximation of $q^*(\alpha|\bar{x})$ for multiple $\bar{x}$, and select the appropriate $q(\alpha|\bar{x})$ based on $\omega_e$ and the vertex on the light source of current $\bar{x}$. The memory cost of the cache is within 16MB.

The equal-sample comparison between SMIS2/SMIS4 and our method is shown in \autoref{fig:photon plane}. Note that SMIS and our method incur different computational costs for evaluating $p(\alpha, \bar{x})$ to estimate the CMIS weight, $\tilde{\omega}(\alpha, \bar{x})$. Our method achieves more efficient CMIS weight estimation with lower computation cost, showing the advantages of our reciprocal estimation method. In terms of SMAPE (Symmetric Mean Absolute Percentage Error) and MSE, our approach also shows superior performance over the alternative approaches.

\section{Conclusion, Limitation, and Future Work} 
In this paper, we have proposed a novel method for tracing difficult specular-involved paths based on efficient reciprocal estimation. By discarding problematic segments of the original path and tracing a proxy path that can meet specular constraint, we enable a probabilistic connection to handle specular paths well. To ensure unbiased estimations, we introduce a sub-optimal RRS function that significantly boosts efficiency. Additionally, we propose an optimal setting of $B$, minimizing the upper bound on the cost of our reciprocal estimation. Specifically, This optimized setting not only aids in the reciprocal estimation required for tracing incomplete sub-paths but also employs a light sub-path reuse strategy to diminish the overhead associated with reciprocal estimations. We have developed a method that allows for replacing small segments of the path, resulting in significant improvements. We believe this approach serves as a powerful tool, enabling the seamless substitution of problematic paths with preferable alternatives while maintaining the unbiased nature of the Monte Carlo estimation of the path integral. 

To provide an intuitive understanding, our method improves connection efficiency by broadening the narrow lobe of the BSDF on specular surfaces to encompass a wider range. Specifically, for the $L_{DD}S^*S$ type light sub-path, this expanded range correlates with the projection area of an area light source at the specular vertex. This enhancement facilitates the inclusion of an eye sub-path within the valid range more easily, particularly when the eye sub-path is far from the specular vertex or when the light source is large. However, challenges arise when the specular surface is far from a small light source or when the diffuse eye vertex is near the specular light vertex. Under such conditions, our connection strategy may struggle to be effective. Despite the assistance of subspace-based probabilistic connections, the efficiency of our method may still face limitations. This is a fundamental challenge of connection-based sampling strategies, highlighting an area ripe for future research endeavors.

Another limitation of our method pertains to the sampling efficiency of proxy vertices, which yields optimal results when these vertices are readily samplable. Constructing an efficient supporting distribution for reciprocal estimation presents significant challenges. To mitigate these issues, we have incorporated path guiding techniques, as demonstrated in the \emph{Water} scene depicted in \autoref{fig:PGCompare}, specifically to address scenarios involving close specular-diffuse interactions. Furthermore, it is feasible to integrate path guiding with our method to develop a more efficient supporting distribution. However, the specifics of such an integration require further detailed exploration.

In addition to enhancing the rendering of paths involving specular interactions, our proposed proxy sampling method holds promise for a range of applications. It allows for the retracing of problematic segments of a full path while preserving most existing vertices, complementing current sampling methodologies. Identifying appropriate scenarios for its application and reducing the computational overhead associated with reciprocal estimation are challenges we aim to tackle in our future research. Furthermore, our method has the potential to become a state-of-the-art BDPT algorithm. It can be effectively integrated with the recent Progressive Photon Mapping (PPM) algorithm to form a highly efficient Vertex Connection and Merging (VCM) technique \cite{Lin2023TVCG}. This integration could significantly speed up the VCM technique. Exploring this integration represents a promising avenue for our future research.

%\begin{acks}
%We extend our gratitude to all the anonymous reviewers for their helpful suggestions. This work is supported by the \grantsponsor{nkpc}{National Key R\&D Program of China}{} (No.~\grantnum{nkpc}{2023YFF0905103}) and~\grantsponsor{nsfc}{NSFC of China}{} (No.~\grantnum{nsfc}{62172013}). 
%We also wish to thank all the test scenes providers: MrChimp2313 (House), Wig42 (Hallway), SlykDrako (Bedroom), Jay-Artist (Kitchen),  Yaoyi Bai et al. (Projector), Benedikt Bitterli (Water).
%We would like to thank Mr. Jierui Ren and Mr. Xiaobai Chen for their assistance in making parts of the illustrations.
%\end{acks}
 
\bibliographystyle{ACM-Reference-Format}
\bibliography{c}

\appendix

\section*{Appendix}
We provide a detailed description of our reciprocal estimator, along with the convergence condition, expectation, variance analysis, and optimal settings.

\section{Our Estimator}
\label{appendix:estimator}

%\begin{CJK*}{UTF8}{gbsn}  
Given:
\begin{itemize}
    \item Target function \qquad \qquad $f(x)>0$
    \item Sample distribution \qquad \  $p(x)\ s.t.\ p(x)>0\ when\ f(x)>0$
    \item Sample \qquad \qquad \qquad \quad \  $X \sim p(x)$
    \item RRS function \qquad \qquad \quad  $r(x|\bar{z})>0$
    \item Constant \qquad \qquad \qquad \ \ $B>0$
\end{itemize}
Here $\bar{z}$ refers to the precedent samples of the estimation.

We give our estimator $\widetilde{I}$, corresponding to $\tilde{I}(\bar{x})$ 
in \autoref{equ:RRS_estimator} of the text, that estimates $\frac{1}{\int _R f(x)dx}$.

\begin{algorithm}
    \caption{Estimator $\widetilde{I}$}
    \label{estimator}
    
    \begin{algorithmic}[1] % The number tells where the line numbering should start
        \Procedure{$\widetilde{I}$}{}
        \State \textbf{return} $\widetilde{I}'(\emptyset )+1/B$
        \EndProcedure
        \Procedure{$\widetilde{I}'(\bar{z})$}{}
            \State $x_0\gets p.sample()$
            \State $g_0\gets 1 - f(x_0)/(B*p(x_0))$
            \State $r_0\gets r(x_0|\bar{z})$
            \State $r_0'\gets \lfloor{r_0}\rfloor$
            \State $ans\gets g_0 / B$
            \If{$uniform(0, 1).sample()$ < $r_0$ -$ r_0'$}
                \State $r_0'\gets r_0'+1$
            \EndIf
            \For{$i = 0$ to $r_0'$}
                \State $ans \gets ans +  g * \widetilde{I}'(\bar{z} + x_0) / r_0$
            \EndFor
            \State \textbf{return} $ans$
        \EndProcedure
    \end{algorithmic}
\end{algorithm}
Note that the target of $\widetilde{I}'(\bar{z})$ is to estimate $\frac{1}{\int_{\mathcal{R}} f(x)dx} - \frac{1}{B}$ whatever the value of the president samples $\bar{z}$ and $\bar{z}$ is irrelevant to $g(x)$. It is obvious that an optimal RRS function $r^*(x|\emptyset)$ for $I'(\emptyset)$ is also optimal for any $I'(\bar{z})$ of president samples $\bar{z}$. Therefore, we simply ignore the existence of $\bar{z}$ and discuss the optimization of an RRS function $r(x)$ that only depends on the current sample $x$ in the rest of our derivation.

To facilitate our derivation, we present another estimator $I$ corresponding to $\widetilde{I}_{sign}$ in the text.
This estimator is almost identical to $\widetilde{I}$, except for the different value added to $ans$. It's feasible to discuss $I$ although our real estimator is $\widetilde{I}$, because:
\begin{enumerate}
    \item Our derivation for ${I}$ also works for $\widetilde{I}$ with an additional approximation.
    \item  $\widetilde{I}$ is always better than $I$, thus decreasing the $I$'s variance will also decrease an upper bound for $\widetilde{I}$'s variance. We will show this in \autoref{apdx:compare_I_It}.
\end{enumerate}

In the following steps, we will first discuss the convergence condition, expectation, and variance of $I$, then discuss the optimal RRS function $r(x)$ and $B$. Finally, we will show $\widetilde{I}$ is always better than $I$, i.e., has less variance.

\begin{algorithm}
    \caption{Estimator I}
    \label{estimator}
    
    \begin{algorithmic}[1] % The number tells where the line numbering should start
        \Procedure{I}{}
            \State $x_0\gets p.sample()$
            \State $g_0\gets 1 - f(x_0)/(B*p(x_0))$
            \State $r_0\gets r(x)$
            \State $r_0'\gets \lfloor{r_0}\rfloor$
            \State $ans\gets 1 / B$
            \If{$uniform(0, 1).sample()$ < $r_0$ -$ r_0'$}
            \State $r_0'\gets r_0'+1$
            \EndIf{}
            
            \For{$i = 0$ to $r_0'$}
                \State $ans \gets ans +  g * I() / r_0$
            
            \EndFor{}
            
            \State \textbf{return} $ans$
        \EndProcedure
    \end{algorithmic}
\end{algorithm}

\section{Convergence Condition}
\label{apdx:Convergence Condition}
Let $g(x) = 1 - \frac{f(x)}{Bp(x)}$ and $\{I_i\}(i\in \mathbb{N})$ be a sequence of independently and identically distributed random variables with respect to $I$, we have
\begin{equation}
\label{equation:I eq1}
I = \frac{1}{B}+\frac{g(X)}{r(X)}\sum\limits_{i}^{r(X)}I_i \ .
\end{equation}
Here $\sum\limits_{i}^{r}I_i = \sum\limits_{i}^{\lfloor{r}\rfloor}I_i+I^*$, with
$$
I^*=\left\{
\begin{array}{rc}
I_{\lceil{r}\rceil} \quad probability\ r-\lfloor{r}\rfloor \\
0\quad \quad probability\ \lceil{r}\rceil-r  \\
\end{array}
\right.
$$

As mentioned in the text, our estimator evaluates a Taylor expansion, namely

$$
\begin{aligned}
    \frac{1}{\int_R f(x)dx} &= \frac{1}{B}\frac{1}{1-E[g(X)]} \\
                            &= \frac{1}{B}(1 + E[g(X)]+E^2[g(X)] + ...) \ .
\end{aligned}
$$

According to the convergence domain of the Taylor expansion, $I$ converges when $E[g(X)]\in(-1,1)$. Since $E[g(X)]=1-\frac{1}{B}\int_Rf(x)dx$, the convergence condition for $I$ is 
\begin{equation}
\label{equation:convergence}
\frac{1}{B}\int_Rf(x)dx\in(0,2)  \ .
\end{equation}

When $E[g(x)]\in(-1,1)$, $E[I]$ is convergent. However, $E[I^2]$ or $V[I]$ are harder to converge and need a stricter condition. We will next prove that in order to let $E[I^2]$ converge, we at least require
\begin{equation}
    E[|g(X)|] < 1  \ .
\end{equation}

We will prove this by contradiction. If $E[|g(x)|]>=1$ and $E[I^2]$ still converges, it holds that 
$$
\begin{aligned}
     E[(I-\frac{1}{B})^2] &= E[\frac{g^2(X)}{r^2(X)} E[(\sum^{r(X)}I_{i})^2]]\\ &\geq E[\frac{g^2(X)}{r^2(X)}{r(X)} E[I ^2]] \\
     &=E[\frac{g^2(X)}{r(X)}] E[I ^2] 
     \\&\geq E[\frac{g^2(X)}{r(X)}]E[(I-\frac{1}{B}) ^2]  \ .
\end{aligned}
$$
Here $E[I^2]\geq E[(I-\frac{1}{B})^2]$, because
$$
E[(I - \frac{1}{B})^2] - E(I^2) =  \frac{1}{B}(\frac{1}{B}-\frac{2}{\int_Rf(x)dx}) \leq 0,
$$
which is assured by the condition 
(\autoref{equation:convergence}).

Now we have 
\begin{equation}
\label{equation:contradiction condition}
E[(I-\frac{1}{B})^2] \geq E[\frac{g^2(X)}{r(X)}]E[(I-\frac{1}{B}) ^2].
\end{equation}
Since $E[|g(X)|]\geq 1$, with Cauchy's inequality, it's easy to prove 
$$
E[\frac{g^2(X)}{r(X)}] > \frac{E^2[|g(X)]|}{E[r(X)]} > 1.
$$
The equality can not hold because $E[r(x)]<1$ is required for a legal RRS function.  This raises the contradiction with 
(\autoref{equation:contradiction condition}) as $E[(I-\frac{1}{B})^2]>0$. Therefore, we conclude that when $E[|g(x)|] \geq 1$, $E[I^2]$ is not convergent, and so is $V[I]$.

\section{Expectation}
The expectation is as:
$$
\begin{aligned}
E[I] &= \frac{1}{B} + E[\frac{g(X)}{r(X)}\sum\limits_{i}^{r(X)}I_i]\\
     &=\frac{1}{B} + E[{g(X)}]E(I)\\
     &= \frac{1}{B(1-E[g(X)])}\\
     &=\frac{1}{\int_Rf(x)dx}  \ .
\end{aligned}
$$

Therefore, we have
\begin{equation}
  E[I]=\frac{1}{\int_Rf(x)dx}   \ .
\end{equation}

\section{Variance}
Recalling previously defined $\sum\limits_{}^{r}I_i$, when $r$ is an integer, it's easy to derive 
\begin{equation}
\label{equation: aprrox1}
E[(\sum\limits_{i}^{r}I_i)^2]=rE[I^2]+(r^2-r)E^2[I]    \ .
\end{equation}

We adopt an approximation\footnote{This approximation is also adopted in \citet{Rath2022}} and use this result when $r$ is a positive real number. With this approximation, we derive the variance of $I$:
$$
\begin{aligned}
V[I] &=V[I-\frac{1}{B}]\\
     &=E[(I-\frac{1}{B})^2]-E^2[I-\frac{1}{B}]\\
     &=E[\frac{g^2(X)}{r^2(X)}(\sum\limits_{i}^{r(X)}I_i)^2] - (E[I]-\frac{1}{B})^2\\
     &\approx E[\frac{g^2(X)}{r(X)}(E[I^2]+(r(X)-1)E^2[I])] - (E[I]-\frac{1}{B})^2\\
     &=E[\frac{g^2(X)}{r(X)}]V[I]+E[g^2(X)]E^2[I] - (E[I]-\frac{1}{B})^2.
\end{aligned}
$$

Let $c_0=E[g^2(x)]E^2[I] - (E[I]-\frac{1}{B})^2$, note that $c_0$ is irrelevant with $r(x)$, then
$$
V[I] = E[\frac{g^2(X)}{r(X)}]V[I]+c_0.
$$
Therefore, with approximation (\autoref{equation: aprrox1}), we have
\begin{equation}
  V[I] = \frac{c_0}{1-E[\frac{g^2(X)}{r(X)}]}.
\end{equation}

\section{Optimal formulation of r(x)}
Next, we will prove that in order to achieve max efficiency, the optimal formulation of $r(x)\equiv |g(x)|$.
We first define efficiency as the product of cost and variance, namely
\begin{equation}
    E_{ff}=\frac{1}{C\times V[I]}.
\end{equation}
In a single iteration of $I$, the expectation of the number of new iterations is $E[r(X)]$. Therefore, $C=\frac{1}{1-E[r(X)]}$, which is the expected number of runs of I. Obviously, we need $C$ to be a finite positive number, so $E[r(x)]\in(0,1)$ is required. Combining the previously obtained variance of I, we have
$$
\begin{aligned}
E_{ff} &=\frac{1}{c_0}(1-E[r(X)])(1-E[\frac{g^2(X)}{r(X)}])\\
    &=\frac{1}{c_0}(1-\int _R r(x)p(x)dx)(1-\int_R\frac{g^2(x)}{r(x)}p(x)dx).    
\end{aligned}
$$
From the integral form of Acz\'{e}l's inequality\footnote{The Acz\'{e}l’s inequality states that if $a_i, b_i (i = 1,2,...,n)$ are positive numbers such that $a^2_1- \sum\limits^{n}_{i=2}a_i^2>0$ or $b^2_1-\sum\limits^{n}_{i=2}b_i^2>0$, then $(a^2_1- \sum\limits^{n}_{i=2}a_i^2)(b^2_1-\sum\limits^{n}_{i=2}b_i^2)\leq (a_1b_1- \sum\limits_{i=2}^na_ib_i )^2$. The equality holds only if $\frac{a_i}{b_i}=\frac{a_1} {b_1}$ for any $i$. More details in \citet{WU20081196}.}, we derive
\begin{equation}
    E_{ff} \leq \frac{1}{c_0}(1-\int_R|g(x)|p(x)dx)^2. 
\end{equation}
The equality holds if and only if $r(x) \equiv |g(x)|$.

In order to achieve this optimality, $E[|g(X)|]\in(0,1)$ is required. This is not a restriction as we have proved in \autoref{apdx:Convergence Condition} that $E[|g(x)|]<1$ is required for $V[I]$'s convergence.

\section{Optimal value for B}
\label{sec:optimal value for B}
We now assume $E[|g(X)|]\in(0,1)$ and $r(x)$ takes the optimal form $|g(x)|$, the following discussion explores the optimal value for $B$.

Since $r(x)\equiv |g(x)|$, now we have
$$
\begin{aligned}
E_{ff} &= \frac{1}{c_0}(1-\int_R|g(x)|p(x)dx)^2\\
    &= \frac{(1-E[|g(X)|])^2}{E[g^2(x)]E^2[I] - (E[I]-\frac{1}{B})^2}\\
    &= \frac{(1-E[|g(X)|])^2}{E[g^2(X)]E^2[I] - E^2[g(X)]E^2[I]}\\
    &= \frac{(1-E[|g(X)|])^2}{V[g(X)]}\times \frac{1}{E^2[I]}.\\
\end{aligned}
$$
In the derivation above, we used a relation between $E[g(X)]$ and $E[I]$ : $E[g(X)]E[I]=E[I]-\frac{1}{B}$. It's easy to get this relation since $E[g(X)] = 1 - \frac{1}{BE[I]}$. 

Note that $g(x) = 1 - \frac{f(x)}{Bp(x)}$, therefore $V[g(X)]=V[g(X)-1]=\frac{1}{B^2}V[\frac{f(X)}{p(X)}]$, thus
$$
E_{ff}={(B-B\times E[|g(X)|])^2}\times \frac{1}{E^2[I]V[\frac{f(X)}{p(X)}]}.
$$
Let $c_1=\frac{1}{E^2[I]V[\frac{f(X)}{p(X)}]}$, note that $c_1$ is irrelevant with $B$, then
$$
\begin{aligned}
 E_{ff} &=c_1 (B - \int _R |B-\frac{f(x)}{p(x)}|p(x)dx)^2\\
     &=c_1 (\int _R(B -|B-\frac{f(x)}{p(x)})|p(x)dx)^2\\
     &\leq c_1 (\int _Rf(x)dx)^2.\\
\end{aligned}
$$
The equality condition is $B\geq max\{\frac{f(x)}{p(x)}\}$. We assume $p(x)$ to be a feasible sampling distribution for $f(x)$, i.e. $max\{\frac{f(x)}{p(x)}\}<+\infty$ so the equality condition can be achieved.

As $B$ becomes greater than $max\{\frac{f(x)}{p(x)}\}$, the efficiency remains constant. However, this conclusion was derived under approximation (\autoref{equation: aprrox1}). When we assume $B\geq max\{\frac{f(x)}{p(x)}\}$, $I$ degenerates into a simple form that is similar to the geometric distribution, and the computation of $V[I]$ no longer needs approximation. We review our derivation under this context.

With $B\geq max\{\frac{f(x)}{p(x)}\}$, $g(x) = 1- \frac{f(x)}{Bp(x)}\in(0,1)$, $I$ degenerates into a simple form:
$$
I = \frac{1}{B}+
\left\{\begin{array}{rc}
I\quad probability\ \ \ \ \ \ g(X) \ \ \ \  \\
0\quad probability\ \ \ 1-g(X)  \\
\end{array}
\right.
$$
Let $q = \int_Rg(x)p(x)dx$, which assembles the continue probability in geometric distribution, then
$$
C = \frac{1}{1-q},\ V[I] =\frac{q}{(B-Bq)^2}=\frac{q}{E^2[I]}.
$$
We have
$$
E_{ff} = \frac{1}{C\times V[I]}=(\frac{1}{q}-1)E^2[I].
$$

Therefore, as $q$ decreases, the efficiency increases. However, $q$ is a monotonically increasing function of $B$. Taking into account the condition $B\geq \max\{\frac{f(x)}{p(x)}\}$, we consider $B=\max\{\frac{f(x)}{p(x)}\}$ as the optimal value.

\section{Comparison between $I$ and $\widetilde{I}$}
\label{apdx:compare_I_It}
While we have been using $I$ for the proof previously, our derivation also apply to $\widetilde{I}$ using an additional approximation, namely
$$
\begin{aligned}
V[\widetilde{I}] &= V[\frac{g(X)}{B} + \frac{g(X)}{r(X)}\sum\limits^{r(x)}\widetilde{I_i}]\\
                 &\approx V[\frac{g(X)}{B}] + V[\frac{g(X)}{r(X)}\sum\limits^{r(x)}\widetilde{I_i}].
\end{aligned}
$$
which assumes the first part and the second part in RHS is irrelevant. With this approximation, the previous conclusion $r(x)\equiv g(x)$
and $B\geq max\{\frac{f(x)}{p(x)}\}$ still holds true, but the last part in \autoref{sec:optimal value for B} that removes approximation (\autoref{equation: aprrox1}) can not be easily applied to $\widetilde{I}$, as $V[\widetilde{I}]$ becomes a complex function of $B$ and simple monotonicity analysis is no longer feasible. But we will prove $B=max\{\frac{f(x)}{p(x)}\}$ is still sub-optimal for $\widetilde{I}$ in the end.

We first show $\widetilde{I}$'s variance is less than $I$. Recall that
$$
I = \frac{1}{B}+\frac{g(X)}{r(X)}\sum\limits_{i}^{r(X)}I_i
$$
and
$$
\widetilde{I'} = \frac{g(X)}{B}+\frac{g(X)}{r(X)}\sum\limits_{i}^{r(X)}\widetilde{I'_i}.
$$
Let $I' = I-\frac{1}{B}$, we have
$$
I' = \frac{g(X)}{r(X)}\sum\limits_{i}^{r(X)}(I'_i+\frac{1}{B}).
$$
Note that $V[I] = V[I']$ and $V[\widetilde{I}]=V[\widetilde{I'}]$, so we can just compare $V[I']$ and $V[\widetilde{I'}]$. In addition, with $E[I']=E[\widetilde{I'}]=\frac{1}{\int_R{f(x)dx}}-\frac{1}{B}$, the comparison between $V[I]$ and $V[\widetilde{I}]$ eventually turns into the comparison between $E[I'^2]$ and $E[\widetilde{I'}^2]$:

For $I$,
$$
E[\widetilde{I'}^2]=E[\frac{g^2(X)}{r^2(X)}(\frac{r(X)}{B}+\sum\limits_{i}^{r(X)}\widetilde{I_i})^2]
$$
and for $I'$,
$$
\begin{aligned}
E[I'^2] &= E[\frac{g^2(X)}{r^2(X)}(\sum\limits_{i}^{r(X)}(I_i+\frac{1}{B}))^2]\\
        &=E[\frac{g^2(X)}{r^2(X)}(\frac{\lfloor r(X) \rfloor}{B}+\frac{\widetilde{r(X)}}{B}+\sum\limits_{i}^{r(X)}I_i)^2]
\end{aligned}
$$
where 
$$
\widetilde{r} = 
\left\{\begin{array}{rc}
1\quad probability\ \ \ r-\lfloor r\rfloor \\
0\quad probability\ \ \ \lceil r\rceil-r  \\
\end{array}
\right.
$$

Let $\widetilde{Z'}=\frac{r}{B}+\sum\limits_{i}^{r}\widetilde{I_i}$ and $Z'=\frac{\lfloor r \rfloor}{B}+\frac{\widetilde{r}}{B}+\sum\limits_{i}^{r}I_i$.
When $r$ is an integer, $E[Z'^2]=E[\widetilde{Z'}^2]$. Otherwise, $\widetilde{r(X)}\neq 0$, we can prove $E[\widetilde{Z'}^2] < E[Z'^2]$, and $E[\widetilde{I'}^2] < E[I'^2]$. Therefore, we eventually prove $V[\widetilde{I}] \leq V[I]$, which means $\widetilde{I}$ is a better estimator.

As we have derived in \autoref{sec:optimal value for B}, $B = max\{\frac{f(x)}{p(x)}\}$ is optimal for $I$.
Although we only attain $B \geq max\{\frac{f(x)}{p(x)}\}$ for $\widetilde{I}$, since $V[I]$ is an upper bound  for $V[\widetilde{I}]$, $B=max\{\frac{f(x)}{p(x)}\}$ can still be considered sub-optimal for $\widetilde{I}$.

\end{document}